\documentclass[a4paper]{article}

\usepackage{caption}
\usepackage{subcaption}

\usepackage[affil-it]{authblk}

\usepackage{longtable}

\usepackage[utf8]{inputenc}
\usepackage{times}
\usepackage{authblk}

\usepackage[]{algorithm, algorithmic}
\newcommand{\setalglineno}[1]{%
  \setcounter{ALC@line}{\numexpr#1-1}}

\usepackage{graphicx}
\usepackage{amsmath}
\usepackage{amsmath, amssymb}
\DeclareMathOperator{\diag}{diag}

\usepackage{booktabs}
\usepackage{color}
\usepackage[usenames,dvipsnames]{xcolor}
\usepackage[round]{natbib}

\usepackage{hyperref}

\usepackage[top=1.25in, bottom=1.25in, left=1.25in, right=1.25in]{geometry}

\DeclareMathOperator*{\argmin}{arg\,min}

\providecommand{\keywords}[1]
{
  \small	
  \textbf{\textit{Keywords---}} #1
}

\makeatletter
\newenvironment{breakablealgorithm}
  {
   \begin{center}
     \refstepcounter{algorithm}
     \hrule height.8pt depth0pt \kern2pt
     \renewcommand{\caption}[2][\relax]{
       {\raggedright\textbf{\ALG@name~\thealgorithm} ##2\par}
       \ifx\relax##1\relax 
         \addcontentsline{loa}{algorithm}{\protect\numberline{\thealgorithm}##2}
       \else
         \addcontentsline{loa}{algorithm}{\protect\numberline{\thealgorithm}##1}
       \fi
       \kern2pt\hrule\kern2pt
     }
  }{
     \kern2pt\hrule\relax
   \end{center}
  }
\makeatother

\usepackage{graphicx}
\usepackage{subcaption}

\usepackage{fancyhdr}
\usepackage{chngcntr}
\counterwithin{figure}{section}
\counterwithin{table}{section}

\usepackage{booktabs}
\usepackage{multirow}

\usepackage{appendix}

\begin{document}
\title{Adaptive Step-Length Selection in Gradient Boosting for Generalized Additive Models for Location, Scale and Shape}
%% Information for the first author.
\author[$1$,*]{Boyao Zhang}
\author[$1$]{Tobias Hepp}
\author[$2$]{Sonja Greven}
\author[$1$]{Elisabeth Bergherr}

\affil[$1$]{Department of Medical Informatics, Biometry and Epidemiology, Friedrich-Alexander Universität Erlangen-Nürnberg, Waldstrasse 6, 91054 Erlangen, Germany}
\affil[$2$]{Chair of Statistics, School of Business and Economics, Humboldt-Universität zu Berlin, Unter den Linden 6, 10099 Berlin, Germany}
\affil[*]{Corresponding author {\sf{e-mail: boyao.zhang@fau.de}}, Phone: +49-(0)9131-85-22729, FAX: +49-(0)9131-85-25740}

\maketitle  

\begin{abstract}
Tuning of model-based boosting algorithms relies mainly on the number of iterations, while the step-length is fixed at a predefined value. 
For complex models with several predictors such as Generalized Additive Models for Location, Scale and Shape (GAMLSS), imbalanced updates of predictors, where some distribution parameters are updated more frequently than others, can be a problem that prevents some submodels to be appropriately fitted within a limited number of boosting iterations. 
We propose an approach using adaptive step-length (ASL) determination within a non-cyclical boosting  algorithm for GAMLSS to prevent such imbalance. Moreover, for the important special case of the Gaussian distribution, we discuss properties of the ASL and derive a  semi-analytical form of the ASL that avoids manual selection of the search interval and numerical optimization to find the optimal step-length, and consequently improves computational efficiency.
We show competitive behavior of the proposed approaches compared to penalized maximum likelihood and boosting with a fixed step-length for GAMLSS models in two simulations and two applications, in particular for cases of large variance and/or more variables than observations.
In addition, the idea of the ASL is also applicable to other models with more than one predictor like zero-inflated count model, and brings up insights into the choice of the reasonable defaults for the step-length in simpler special case of (Gaussian) additive models.
\end{abstract}


\normalsize
\section{Introduction}
Generalized additive models for location, scale and shape (GAMLSS)[\cite{doi:10.1111/j.1467-9876.2005.00510.x}] are distribution-based approaches, where all parameters of the assumed distribution for the response can be modelled as additive functions of the explanatory variables [\cite{doi:10.1142/9781860945410_0007, stasinopoulos2017}].
Specifically, the GAMLSS framework allows the conditional distribution of the response variable to come from a wide variety of discrete, continuous and mixed discrete-continuous distributions, see \cite{JSSv023i07}. 
Unlike conventional generalized additive models (GAMs), GAMLSS
not only model the location parameter, e.g. the mean for Gaussian distributions, but also further distribution parameters such as scale (variance) and shape (skewness and kurtosis) through the explanatory variables in linear, nonlinear or smooth functional form.

The coefficients of GAMLSS are usually estimated based on penalized maximum likelihood method [\cite{doi:10.1111/j.1467-9876.2005.00510.x}]. 
However, this approach cannot deal with high dimensional data, or more precisely, the case of more variables than observations [\cite{buehlmann2006}]. 
As the selection of informative covariates is an important part of practical analysis, \cite{Mayr} combined the GAMLSS framework with componentwise gradient boosting [\cite{doi:10.1198/016214503000125, Hofner:2014:MBR:2582354.2582382, mboost}] such that variable selection and estimation can be performed simultaneously. 
The original method cyclically updates the distribution parameters, i.e. all predictors will be updated sequentially in each boosting iteration [\cite{JSSv074i01}]. 
Because the levels of complexity vary across the prediction functions, separate stopping values are required for each distribution parameter. 
Consequently, these stopping values have to be optimized jointly as they are not independent of each other. 
The commonly applied joint optimization methods like grid search are, however, computationally very demanding. 
For this reason, \cite{Thomas2018} proposed an alternative non-cyclical algorithm that updates only one distribution parameter (yielding the strongest improvement) in each boosting iteration. 
This way, only one global stopping value is needed and the resulting one-dimensional optimization procedure vastly reduces computing complexity for the boosting algorithm compared to the previous multi-dimensional one. 
The non-cyclical algorithm can be combined with stability selection [\cite{doi:10.1111/j.1467-9868.2010.00740.x, Hofner2015}] to further reduce the selection of false positives [\cite{mboost2}].

In contrast to the cyclical approach, the non-cyclical algorithm avoids an equal number of updates for all distribution parameters as it is not useful to artificially enforce updates for parameters with a less complex structure than other parameters.
However, it becomes even more important to fairly select the predictor to be updated in any given iteration.
%The non-cyclical algorithm avoids the equal number of updates for all distribution parameters that the cyclical approach uses. 
%On the one hand, this is reasonable, as it is not useful to artificially enforce updates for parameters that have a less complex structure than other parameters. 
%On the other hand, this means that a fair selection of the  predictor that is updated in any given iteration becomes all the more important. 
The current implementation of \cite{Thomas2018}, however, uses fixed and equal step-lengths for all updates, regardless of the achieved loss reduction of different distribution parameters.
As we demonstrate later, this leads to imbalanced updates that affect the fair selection and predictors with large number of boosting iterations still tend to be underfitted.
%For the Gaussian distribution, this happens especially in scenarios with relatively large variance, as this decreases the negative gradient for the loss of the mean and consequently reduces the selection chance of corresponding base-learners in each iteration.
%Moreover, even if a base-learner for the mean is selected, the improvement will be very small and thus a lot of updates are required.
This seems inconsistent, since one expects the underfitted predictor to be updated with a few number of iterations.
As we show later, a large $\boldsymbol{\sigma}$ in a Gaussian distribution leads to a small negative gradient of $\boldsymbol{\mu}$ and consequently the improvement for $\boldsymbol{\mu}$ with fixed small step-lengths in each boosting iteration will also be small.
This results in the algorithm needing a lot of updates for $\boldsymbol{\mu}$ until its empirical risk decreases to the level of $\boldsymbol{\sigma}$. 
However, the algorithm may stop long before the corresponding coefficients are well estimated. %A possible solution to the problem is to increase the step length for $\mu$ in each boosting iteration, but choosing an appropriate value is not trivial, because a large step length will lead to overfitting, while a small one does not help a lot.

We address this problem by proposing a version of the non-cyclical boosting algorithm for GAMLSS that  adaptively and automatically optimizes the step-lengths for all predictors in each boosting iteration. 
The new approach leads to a fair selection of predictors to update. 
While it does not enforce equal numbers of updates for all distribution parameters, it yields a natural balance in the updates. 
For the Gaussian distribution, we also derive  (semi-)analytical adaptive step-lengths that decrease the need for numerical optimization and discuss their properties. Our findings have implications beyond boosted GAMLSS models for boosting other models with several predictors, e.g. for zero-inflated count models, and also give insights into the step-length choice for the simpler special case of (Gaussian) additive models.

The structure of this paper is organized as follows: 
Section \ref{sec: boost_gamlss} introduces the boosted GAMLSS models including the cyclical and non-cyclical algorithms. 
Section \ref{sec: asl} discusses how to apply the adaptive step-length on the non-cyclical boosted GAMLSS algorithm, and introduces the semi-analytical solutions of the adaptive step-length in the Gaussian distribution with their properties. 
Section \ref{sec: sim} evaluates the performance of the adaptive algorithms and the problem of fixed step-length in two  simulations. 
Section \ref{sec: empirical} presents the application of the adaptive algorithms for two datasets: the malnutrition data, where the outcome variance is very large, and the riboflavin data, which has more variables than observations. 
Section \ref{sec: conclusion} concludes with a summary and discussion. 
Further relevant materials and results are included in the appendix.

\section{Boosted GAMLSS}
In this section, we briefly introduce the GAMLSS models and the two cyclical and noncyclical boosting methods for estimation. 
\label{sec: boost_gamlss}
\subsection{GAMLSS and componentwise gradient boosting}
Conventional generalized additive models (GAM) assume a dependence only of the conditional mean $\mu$ of the response on the covariates. 
GAMLSS, however, also model other distribution parameters such as the scale $\sigma$, skewness $\nu$ and/or kurtosis $\tau$ with a set of statistical models. %Theoretically, the framework of GAMLSS allows fitting the distribution with infinite parameters.

%Conventional generalized additive models (GAM) assumes only the dependency between the response and the conditional mean. GAMLSS, however, regresses the distribution of the univariate response with a set of statistical models. Apart from the conditional mean or location parameter $\mu$, the scale $\sigma$, skewness $\nu$ and kurtosis $\tau$ parameter are the common used distribution parameters. But theoretically, the framework of the GAMLSS allows to fit the distribution with infinite parameters.

The $K$ distribution parameters $\boldsymbol{\theta}^T = (\boldsymbol{\theta}_1, \boldsymbol{\theta}_2, \cdots, \boldsymbol{\theta}_K)$ of a density function $f(\boldsymbol{y}|\boldsymbol{\theta})$ are modelled by a set of up to $K$ additive models. 
The model class assumes that the observations $y_i$ for $i \in \{1, \cdots, n\}$ are conditionally independent given a set of explanatory variables. 
Let $\boldsymbol{y}^T = (y_1, y_2, \cdots, y_n)$ be a vector of the response variable and $\boldsymbol{X}$ be a $n \times J$ data matrix. 
In addition, we denote $\boldsymbol{X}_{i \cdot}$, $\boldsymbol{X}_{\cdot j}$ and $X_{ij}$ as the $i$-th observation (vector of length $J$), $j$-variable (vector of length $n$) and the $i$-th observation of the $j$-th variable (a single value) respectively. 
Let $g_k(\cdot), k = 1, \cdots, K$ be known monotonic link functions that relate $K$ distribution parameters to explanatory variables through additive models given by
\begin{align}
\label{gamlss}
g_k(\boldsymbol{\theta}_k) = \eta_{\boldsymbol{\theta}_k}(\boldsymbol{X}) = \beta_{0, \boldsymbol{\theta}_k} \boldsymbol{1}_n + \sum_{j=1}^{J}f_{j, \boldsymbol{\theta}_k}(\boldsymbol{X}_{\cdot j} | \beta_{j, \boldsymbol{\theta}_k}) \quad k=1, \dots, K,
\end{align}
where $\boldsymbol{\theta}_k = (\theta_{k, 1}, \cdots, \theta_{k, n})^T$ contains the $n$ parameter values for the $n$ observations and functions are applied elementwise if the argument is a vector, 
$\boldsymbol{\eta}_{\theta_k}$ is a vector of length $n$, $\boldsymbol{1}_n$ is a vector of ones and $\beta_{0, \boldsymbol{\theta}_k}$ is the model parameter specific intercept. 
Function $f_{j, \boldsymbol{\theta}_k}(\boldsymbol{X}_{\cdot j} | \beta_{j, \boldsymbol{\theta}_k})$ indicates the effects of the $j$-th explanatory variable $\boldsymbol{X}_{\cdot j}$ (vector of length $n$) for the model parameter $\boldsymbol{\theta}_k$, and $\beta_{j, \boldsymbol{\theta_k}}$ is the parameter of the additive predictor $f_{j, \boldsymbol{\theta}_k}(\cdot)$. Various types of effects (e.g., linear, smooth, random) for $f(\cdot)$ are allowed.
If the location parameter $(\theta_1 = \mu)$ is the only distribution parameter to be regressed ($K=1$) and the response variable is from the exponential family, \eqref{gamlss} reduces to the conventional GAM.
In addition, $f_j$ can depend on more than one variable (interaction), in which case $X_{\cdot j}$ would be e.g. a $n \times 2$ matrix, but for simplicity we ignore this case in the notation.

A penalized likelihood approach can be used to estimate the unknown quantities; for more details, see \cite{doi:10.1111/j.1467-9876.2005.00510.x}. 
However, this approach does not allow parameter estimation in the case of more explanatory variables than observations, and variable selection for high-dimensional data is not possible. 
To deal with these problems, \cite{Mayr} proposed a boosted GAMLSS algorithm, which estimates the predictors in GAMLSS with componentwise gradient boosting [\cite{doi:10.1198/016214503000125}]. 
As this method updates in general only one variable in each iteration, it can deal with data that has more variables than observations, and the important variables can be selected by controlling the stopping iterations.

To estimate the unknown predictor parameters $\beta_{j, \boldsymbol{\theta}_k}, j \in \{1, \cdots, J\}$ in equation \eqref{gamlss}, the componentwise gradient boosting algorithm minimizes the empirical risk $R$, which is also the loss $\rho$ summed over all observations,
\begin{align*}
R = \sum_{i=1}^n \rho \left(y_i, \boldsymbol{\eta}(\boldsymbol{X}_{i \cdot})\right),
\end{align*}
where the loss $\rho$ measures the discrepancy between the response $y_i$ and the predictor $\boldsymbol{\eta}(\boldsymbol{X}_{i \cdot})$.
The predictor $\boldsymbol{\eta}(\boldsymbol{X}_{i \cdot}) = \left(\eta_{\boldsymbol{\theta}_1}(\boldsymbol{X}_{i\cdot}), \cdots, \eta_{\boldsymbol{\theta}_K}(\boldsymbol{X}_{i\cdot}) \right)$ is a vector of length $K$.
For the $i$-th observation $\boldsymbol{X}_{i\cdot}$, each predictor $\eta_{\boldsymbol{\theta}_k}(\boldsymbol{X}_{i\cdot})$ is a single value corresponding to the $i$-th entry in $\eta_{\boldsymbol{\theta_k}}$ in equation \eqref{gamlss}.
The loss function $\rho$ usually used in GAMLSS is the negative log-likelihood of the assumed distribution of $\boldsymbol{y}$ [\cite{Thomas2018, friedman2000}].

The main idea of gradient boosting is to fit simple regression base-learners $h_j(\cdot)$ to the pseudo-residuals vector $\boldsymbol{u}^T = (u_1, \cdots, u_n)$, which is defined as the negative partial derivatives of loss $\rho$, i.e.,
\begin{align*}
\boldsymbol{u}_k^{[m]} =  \left(-\frac{\partial}{\partial \eta_{\boldsymbol{\theta}_k}}\rho(y, \boldsymbol{\eta}) \Big\vert_{\boldsymbol{\eta} = \hat{\boldsymbol{\eta}}^{[m-1]}(\boldsymbol{X}_{i \cdot}), y = y_i} \right)_{i=1, \cdots, n},
\end{align*}
where $m$ denotes the current boosting iteration. In a componentwise gradient boosting iteration, each base-learner involves usually one explanatory variable (interactions are also allowed) and is fitted separately to $\boldsymbol{u}_k^{[m]}$,
\begin{align*}
\boldsymbol{u}_k^{[m]} \overset{\text{base-learner}}{\longrightarrow} \hat{h}_{j, \boldsymbol{\theta}_k}^{[m]}(\boldsymbol{X}_{\cdot j}) \quad \text{for} \quad j = 1, \cdots, J.
\end{align*}
For linear base-learner, its correspondence to the model terms in \eqref{gamlss} shall be
\begin{align*}
\hat{h}_{j, \boldsymbol{\theta}_k}(\boldsymbol{X}_{\cdot j}) = \boldsymbol{X}_{\cdot j} \hat{\boldsymbol{\beta}}_j,
\end{align*}
where the estimated coefficients can be obtained by using the maximum likelihood or least square method. The best-fitting base-learner is selected based on the residual sum of squares, i.e.,
\begin{align*}
j^* = \argmin_{j \in \{1, \cdots, J\}} \sum_{i=1}^n \left(u_{k, i} - \hat{h}_j(X_{ij})\right)^2,
\end{align*}
thereby allowing for easy interpretability of the estimated model and also the use of hypothesis tests for single base-learners [\cite{hepp2019}].
The additive predictor will be updated based on the best-fitting base-learner $\hat{h}_{j^*, \theta_{k^*}}(\boldsymbol{X}_{\cdot j^*})$ in terms of the best-performing sub-model $\eta_{\boldsymbol{\theta}_{k^*}}$,
\begin{align}
\label{eq: update_eta}
\hat{\eta}_{\boldsymbol{\theta}_{k^*}}^{[m]}(\boldsymbol{X}) = \hat{\eta}_{\boldsymbol{\theta}_{k^*}}^{[m-1]}(\boldsymbol{X}) + \nu \hat{h}_{j^*, \boldsymbol{\theta}_{k^*}}(\boldsymbol{X}_{\cdot j^*}),
\end{align}
where $\nu$ denotes the step-length. 
In order to prevent overfitting, the step-length is usually set to a small value, in most cases 0.1.
Equation \eqref{eq: update_eta} updates only the best-performing predictor $\hat{\eta}_{\boldsymbol{\theta}_{k^*}}^{[m]}$, all other predictors (i.e. for $k \neq k^*$) remain the same as in the previous boosting iteration.
The best-performing sub-model $\boldsymbol{\theta}_{k^*}$ can be selected by comparing the empirical risk, i.e. which model parameter achieves the largest model improvement.

The main tuning parameter in this procedure, as in other boosting algorithms, is how many iterations should be performed before it stops, which is denoted as $m_{\theta_{\text{stop}}}$.
As too large or small $m_{\theta_{\text{stop}}}$ leads to over-/underfitting model, cross-validation [\cite{kohavi1995study}] is one of the most widely used methods to find the optimal $m_{\theta_{\text{stop}}}$.

\subsection{Cyclical boosted GAMLSS}
The boosted GAMLSS can deal with data that has more variables than observations, as the componentwise gradient boosting updates only one variable in each iteration. 
It leads to variable selection if some less important variables have never been selected as the best-performing variable and thus are not included in the final model for a given stopping iteration $m_{\theta_{\text{stop}}}$.

The original framework of boosted GAMLSS proposed by \cite{Mayr} is a cyclical approach, which means every predictor $\eta_{\boldsymbol{\theta}_k}, k \in \{1, \cdots, K\}$ is updated in a cyclical manner inside each boosting iteration.
The iteration starts by updating the predictor for the location parameter and uses the predictors from the previous iteration for all other parameters.
Then, the updated location model will be used for updating the scale model and so on.
%When update one predictor, all other parameters stay fixed, and the updated predictor will be involved in the update of the next predictor.
A schematic overview of the updating process in iteration $m+1$ for $K = 4$ is
\begin{align*}
(\hat{\boldsymbol{\mu}}^{[m]}, \hat{\boldsymbol{\sigma}}^{[m]}, \hat{\boldsymbol{\nu}}^{[m]}, \hat{\boldsymbol{\tau}}^{[m]}) \overset{\text{update}}{\longrightarrow} \hat{\eta}_{\boldsymbol{\mu}}^{[m+1]} \rightarrow \hat{\boldsymbol{\mu}}^{[m+1]} \\
(\hat{\boldsymbol{\mu}}^{[m+1]}, \hat{\boldsymbol{\sigma}}^{[m]}, \hat{\boldsymbol{\nu}}^{[m]}, \hat{\boldsymbol{\tau}}^{[m]}) \overset{\text{update}}{\longrightarrow} \hat{\eta}_{\boldsymbol{\sigma}}^{[m+1]} \rightarrow \hat{\boldsymbol{\sigma}}^{[m+1]} \\
(\hat{\boldsymbol{\mu}}^{[m+1]}, \hat{\boldsymbol{\sigma}}^{[m+1]}, \hat{\boldsymbol{\nu}}^{[m]}, \hat{\boldsymbol{\tau}}^{[m]}) \overset{\text{update}}{\longrightarrow} \hat{\eta}_{\boldsymbol{\nu}}^{[m+1]} \rightarrow \hat{\boldsymbol{\nu}}^{[m+1]} \\
(\hat{\boldsymbol{\mu}}^{[m+1]}, \hat{\boldsymbol{\sigma}}^{[m+1]}, \hat{\boldsymbol{\nu}}^{[m+1]}, \hat{\boldsymbol{\tau}}^{[m]}) \overset{\text{update}}{\longrightarrow} \hat{\eta}_{\boldsymbol{\tau}}^{[m+1]} \rightarrow \hat{\boldsymbol{\tau}}^{[m+1]}.
\end{align*}

However, not all of the distribution parameters have the same complexity, i.e., the stopping iterations $m_{\theta_{\text{stop}}}$ should be set separately for different parameters, or jointly optimized, for example by grid search. 
Since grid search scales exponentially with the number of distribution parameters, such optimization can be very slow.

\subsection{Non-cyclical boosted GAMLSS}
In order to deal with the issues of a cyclical approach, \cite{Thomas2018} proposed a \textit{non-cyclical} version, that updates only one distribution parameter instead of successively updating all parameters in each boosting iteration by comparing the model improvement (negative log-likelihood) of each model parameter, see Algorithm \ref{alg: ncgb} (especially step \ref{alg: ncgb10}).
Consequently, instead of specifying separate stopping iterations $m_{\boldsymbol{\theta}\text{stop}}$ for different parameters and tuning them with the computationally demanding grid search, only one overall stopping iteration, denoted as $m_{\text{stop}}$, needs to be tuned with e.g. the line search [\cite{friedman2001, brent2013algorithms}].
The tuning problem thus reduces from a multi-dimensional to a one-dimensional problem, which vastly reduces the computing time.

\begin{breakablealgorithm}
\caption{Non-cyclical componentwise gradient boosting in multiple dimensions - Basic algorithm}
\label{alg: ncgb}
\begin{algorithmic}[1]
\STATE Initialize the additive predictors $\hat{\boldsymbol{\eta}}^{[0]} = \left(\hat{\eta}_{\boldsymbol{\theta}_1}^{[0]}, \cdots, \hat{\eta}_{\boldsymbol{\theta}_K}^{[0]}\right)$ with offset values.
\STATE For each distribution parameter $\boldsymbol{\theta}_k, k=1, \cdots, K$, specify a set of base-learners, i.e., for parameter $\boldsymbol{\theta}_k$ define $h_{1, \boldsymbol{\theta}_k}(\cdot), \cdots, h_{J_k, \boldsymbol{\theta}_k}(\cdot)$ where $J_k$ is the cardinality of the set of base-learners specified for $\boldsymbol{\theta}_k$.
\FOR {$m=1$ to $m_{\text{stop}}$}
	\FOR {$k=1$ to $K$}
	\STATE Compute negative partial derivatives $-\frac{\partial}{\partial \eta_{\boldsymbol{\theta}_k}}\rho(y, \boldsymbol{\eta})$ and plug in the current estimates $\hat{\boldsymbol{\eta}}^{[m-1]}(\cdot)$:
	\begin{align*}
	\boldsymbol{u}_k^{[m]} = \left(-\frac{\partial}{\partial \eta_{\boldsymbol{\theta}_k}}\rho(y, \boldsymbol{\eta}) \Big\vert_{\boldsymbol{\eta} = \hat{\boldsymbol{\eta}}^{[m-1]}(\boldsymbol{X}_{i\cdot}), y=y_i}\right)_{i=1, \cdots, n}.
	\end{align*}
	\STATE Fit (e.g. with the least square method) the negative gradient vector $\boldsymbol{u}_k^{[m]}$ separately to every base-learner:
		\begin{align*}
		\boldsymbol{u}_k^{[m]} \overset{\text{base-learner}}{\longrightarrow} \hat{h}_{j, \boldsymbol{\theta}_k}(\boldsymbol{X}_{\cdot j}) \quad \text{for} \quad j = 1, \cdots, J_k.
		\end{align*}
	\STATE Select the best-fitting base-learner $\hat{h}_{j^*, \boldsymbol{\theta}_k}(\boldsymbol{X}_{\cdot j^*})$ by inner loss, i.e., the residual sum of squares of the base-learner fit w.r.t. $\boldsymbol{u}_k^{[m]} = \left(u_{k, 1}^{[m]}, \cdot, u_{k, n}^{[m]}\right)^T$:
		\begin{align*}
		j^* = \argmin_{j \in \{1, \cdots, J_k\}} \sum_{i=1}^n \left(u_{k, i}^{[m]} - \hat{h}_{j, \boldsymbol{\theta}_k}(X_{ij})\right)^2,
		\end{align*}
		where we dropped the dependence of $j^*$ on $k$ in the notation for simplicity.
	\STATE Set the step-length to a fixed value $\nu_0$, usually $\nu_0 = 0.1$: \label{alg_step: sl}
		\begin{align*}
		\nu_{\boldsymbol{\theta}_k}^{[m]} = \nu_0
		\end{align*}
	\STATE Compute the possible improvement of this update regarding the outer loss
	\begin{align*}
	\Delta \rho_k = \sum_{i=1}^n \rho \left(y_i, \hat{\eta}_{\boldsymbol{\theta}_k}^{[m-1]}(\boldsymbol{X}_{i \cdot}) + \nu_{\boldsymbol{\theta}_k}^{[m]} \cdot \hat{h}_{j^*, \boldsymbol{\theta}_k}(X_{ij^*}) \right).
	\end{align*}
	\ENDFOR
\STATE Update, depending on the value of the loss reduction,  only the overall best-fitting base-learner $k^* = \argmin_{k \in \{1, \cdots, K\}}\Delta \rho_k$:
\begin{align*}
\hat{\eta}_{\boldsymbol{\theta}_{k^*}}^{[m]}(\boldsymbol{X}) = \hat{\eta}_{\boldsymbol{\theta}_{k^*}}^{[m-1]}(\boldsymbol{X}) + \nu_{\boldsymbol{\theta}_k}^{[m]} \cdot \hat{h}_{j^*, \boldsymbol{\theta}_{k^*}}(\boldsymbol{X}_{\cdot j^*}).
\end{align*} \label{alg: ncgb10}
\STATE Set $\hat{\boldsymbol{\eta}}_{\boldsymbol{\theta}_{k}}^{[m]} := \hat{\boldsymbol{\eta}}_{\boldsymbol{\theta}_{k}}^{[m-1]}$ for all $k \neq k^*$.
\ENDFOR
\end{algorithmic}
\end{breakablealgorithm}

%The non-cyclical boosted GAMLSS algorithm in each iteration selects not only the best-fitting variable for all predictors, but also the best-performing sub-model (e.g.\ scale, location or shape). 
%The other advantages of the cyclical approach compared to penalized maximum likelihood are also inherited by the non-cyclical approach.

The cyclical approach led to an inherent but somewhat artificial balance between the distribution parameters, as the predictors for all distribution parameters are updated in each iteration. 
%in the cyclical approach no longer holds in the non-cyclical approach. Such balance is an artificial balance, and in fact not a must while estimating. On one hand, 
The different final stopping values $m_{\boldsymbol{\theta}\text{stop}}$ for the different distribution parameters - chosen by tuning methods such as cross-validation - allow to stop updates at different times for distribution parameters of different complexity to avoid overfitting.
%and these $m_{\theta\text{stop}}$ are not of the same. On the other hand, given the same $m_{\theta\text{stop}}$ for all parameters, if one sub-model is already well estimated, there is no need to update it any more, as further updates might lead to overfitting for some of the parameters. 
%However, the ``artificial" balance in the cyclical algorithm makes all parameters be updated in each iteration without worrying about a final model that has an only location or scale sub-model. 
%But such an imbalanced situation can happen 
In the non-cyclical algorithm, especially when $m_{\text{stop}}$ is not large enough, there is the danger of an imbalance between predictors. 
If the selection between predictors to update is not fair, this could lead to iterations primarily updating some of the predictors and underfitting others.  
%For example in Gaussian distribution, a large $\sigma$ will make the algorithm spend lots of its computing resources in updating $\mu$ until the empirical risk of $\mu$ decreased to the level that $\sigma$ has. In the meanwhile, if not enough observations are available, $m_\text{stop}$ tuned by CV can be so small that the final model includes only the $\sigma$ sub-model, 
We will provide a detailed example for the Gaussian distribution with large $\boldsymbol{\sigma}$  in Section \ref{sec: sim2}.

A related challenge is to choose an appropriate step-length $\nu_{\boldsymbol{\theta}_k}^{[m]}$ 
for both the cyclical and non-cyclical approaches. 
Tuning the parameters when boosting GAMLSS models relies mainly on the number of boosting iterations ($m_{\text{stop}}$), with the step-length $\nu$ usually set to a small value such as  0.1. 
\cite{buehlmann2007} argued that using a small  step-length like 0.1 (potentially resulting in a larger number of iterations $m_{\text{stop}}$)  had a similar computing speed as using an adaptive step-length performed by doing a line search, but meant an easier tuning task for one parameter ($m_{\text{stop}}$) instead of two. 
%So researchers can focus on the other challenges instead of tuning step length as long as it is a small value. 
However, this result referred to models with a single predictor. 
A fixed step-length can lead to an imbalance in the case of several predictors that may live on quite different scales. 
For example, 0.1 may be too small for $\mu$ but large for $\sigma$. We will discuss such cases analytically  and with empirical evidence in the later sections. 
Moreover, varying the step-lengths for the different sub-models directly influences the choice of the best performing sub-model in the non-cyclical boosting algorithm, thus choosing a subjective step-length is not  appropriate. 
In the following, we denote a fixed predefined step-length such as 0.1 as the \textit{fixed step-length} (FSL) approach.

To overcome the problems stated above, we propose  using an \textit{adaptive step-length} (ASL) while boosting. 
In particular, we propose to optimize the step-length for each predictor in each iteration to obtain a fair comparison between the predictors. 
While the adaptive step-length has been used before, the proposal to use different ASLs for different predictors is new and we will see that this leads to balanced updates of the different predictors.

\section{Adaptive Step-Length}
\label{sec: asl}
In this section, we first introduce the general idea of the implementation of adaptive step-lengths for different predictors to GAMLSS.
For the important special case of a Gaussian distribution with two model parameters ($\mu$ and $\sigma$), we will derive and discuss their adaptive step-lengths and properties, which also serves as an important illustration of the relevant issues more generally.

\subsection{Boosted GAMLSS with adaptive step-length}
Unlike the step-length in equation \eqref{eq: update_eta} and Algorithm \ref{alg: ncgb}, step \ref{alg: ncgb10}, the adaptive step-length may also vary in different boosting iterations according to the loss reduction.

The adaptive step-length can be derived by solving the optimization problem
\begin{align}
\label{eq: asl}
\nu_{j^*, \boldsymbol{\theta}_k}^{*[m]} = \argmin_{\nu} \sum_{i=1}^n \rho\left(y_i, \hat{\eta}_{\boldsymbol{\theta}_k}^{[m-1]}(\boldsymbol{X}_{i \cdot}) + \nu \cdot \hat{h}_{j^*, \boldsymbol{\theta}_k}(X_{ij^*}) \right),
\end{align}
note that $\nu_{j^*, \boldsymbol{\theta}_k}^{*[m]}$ is the \textit{optimal step-length} of the model parameter $\theta_k$ dependent on $j^*$ in iteration $m$.
The optimal step-length is a value that leads to the largest decrease possible of the empirical risk and usually leads to overfitting of the corresponding variable if no shrinkage is used [\cite{Hepp2016ApproachesTR}]. 
Therefore the actual adaptive step-length (ASL) we apply in the boosting algorithm is the product of two parts, the shrinkage parameter $\lambda$ and the optimal step-length $\nu_{j^*, \boldsymbol{\theta}_k}^{*[m]}$, i.e.,
\begin{align*}
\nu_{j^*, \boldsymbol{\theta}_k}^{[m]} = \lambda \cdot \nu_{j^*, \boldsymbol{\theta}_k}^{*[m]}.
\end{align*}
In this article, we take $\lambda = 0.1$, thus 10\% of the optimal step-length. 
By comparison, the fixed step-length $\nu = 0.1$ would correspond to a combination of a shrinkage parameter $\lambda = 0.1$ with the ``optimal" step-length $\nu^*$ set to one.

The non-cyclical algorithm with ASL can be improved by replacing the fixed step-length in step \ref{alg_step: sl} of algorithm \ref{alg: ncgb} with the adaptive one. We formulate this change in Algorithm \ref{alg: ncgbasl}.

\begin{algorithm}
\caption{Non-cyclical componentwise gradient boosting with adaptive step-length - Extension of basic algorithm \ref{alg: ncgb}}
\label{alg: ncgbasl}
\begin{algorithmic}[1]
\REQUIRE $\cdots$ Steps 1-7 equal to algorithm \ref{alg: ncgb} $\cdots$, in addition, choose shrinkage parameter $\lambda$.
\setalglineno{8}
\STATE Find the optimal step-length $\nu_{\boldsymbol{\theta}_k}^{[m]}$ by optimizing the outer loss:
		\begin{align*}
		\nu_{j^*, \boldsymbol{\theta}_k}^{*[m]} = \argmin_{\nu} \sum_{i=1}^n \rho\left(y_i, \hat{\eta}_{\boldsymbol{\theta}_k}^{[m-1]}(\boldsymbol{X}_{i \cdot}) + \nu \cdot \hat{h}_{j^*, \boldsymbol{\theta}_k}(X_{ij^*}) \right),
		\end{align*}
	and set adaptive step-length $\nu_{j^*, \boldsymbol{\theta}_k}^{[m]}$ as the optimal value with shrinkage $\lambda$:
		\begin{align*}
		\nu_{j^*, \boldsymbol{\theta}_k}^{[m]} = \lambda \cdot \nu_{j^*, \boldsymbol{\theta}_k}^{*[m]}.
		\end{align*}
\REQUIRE $\cdots$ Steps 9-13 equal to those in algorithm \ref{alg: ncgb} $\cdots$
\end{algorithmic}
\end{algorithm}

As the parameters in GAMLSS may have quite different scales, updates with fixed step-length can lead to an imbalance between model parameters, especially when $m_{\text{stop}}$ is not large enough. 
When using FSL, a single update for predictor $\eta_{\boldsymbol{\theta}_1}$ may achieve the same amount of global loss reduction than several updates of another predictor $\eta_{\boldsymbol{\theta}_2}$ even if the actually possible contribution of the competing base-learners is similar, because for different scales the loss reductions of $\eta_{\boldsymbol{\theta}_2}$ in these iterations are always smaller than that of $\eta_{\boldsymbol{\theta}_1}$.
%The updates for $\eta_{\theta_2}$ thus in these iterations are nothing more than reducing its empirical risk to a comparable level that $\eta_{\theta_1}$ already has. 
However, such unfair selections can be avoided by using ASL, because the model improvement depends on the largest decrease possible of each predictor, i.e., the potential reduction in the empirical risks of all predictors are on the same level and their comparison therefore is fair. 
Fair selection does not enforce an equal number of updates as in the cyclical approach.
The ASL approach can lead to imbalanced updates of predictors, but such imbalance actually reveals the intrinsically different complexities of each sub-model.

The main contribution of this paper is the proposal to use ASLs for each predictor in GAMLSS. 
This idea can also be applied to other complex models (e.g. zero-inflated count models) with several predictors for the different parameters, because these models meet the same problem, i.e. the scale of these parameters might differ considerably.
If a boosting algorithm is preferred for estimation of such a model, we provide a new solution to address these kinds of problems, i.e. separate adaptive step-lengths for each distribution parameter.

\subsection{Gaussian distribution specification}
In general, the adaptive step-length $\nu$ can be found by optimizing procedures such as a line search.
However, such methods do not help to reveal the properties of adaptive step-lengths and its relationship with model parameters.
Moreover, a line search method searches for the optimal value from a predefined search interval, which can be difficult to find out since too narrow intervals might not include the optimal value and too large intervals increase the searching time.
% Furthermore, the optimal value found by line search is an asymptotic value, which is not as precise as the analytical solution, if the latter exists.
The direct computation from an analytical expression is faster than a search.
By investigating the important special case of a Gaussian distribution with two parameters, we will learn a lot about the adaptive step-length for the general case.

Consider the data points $(y_i, \boldsymbol{x}_{i\cdot}), i \in \{1, \cdots, n\}$, where $\boldsymbol{x}$ is a $n \times J$ matrix. 
Assume the true data generating mechanism is the normal model
\begin{align*}
y_i &\sim N(\mu_i, \sigma_i) \\
\mu_i &= \eta_{\boldsymbol{\mu}}(\boldsymbol{x}_{i \cdot})\\
\sigma_i &= \exp\left(\eta_{\boldsymbol{\sigma}}(\boldsymbol{x}_{i \cdot})\right).
\end{align*}
As we talk about the observed data, we replace $\eta_{\boldsymbol{\theta}_k}$, where $k = 1, 2$ for Gaussian distribution, with $\boldsymbol{\mu}$ and $\boldsymbol{\sigma}$, and replace $\boldsymbol{X}$ with $\boldsymbol{x}$.
The identity and exponential functions for $\boldsymbol{\mu}$ and $\boldsymbol{\sigma}$ are thus the corresponding inverse link.
Taking the negative log-likelihood as the loss function, its negative partial derivatives $\boldsymbol{u}_{\boldsymbol{\mu}}$ and $\boldsymbol{u}_{\boldsymbol{\sigma}}$ in iteration $m$ for both parameters can then be modelled with the base-learners $\hat{h}_{j, \boldsymbol{\mu}}^{[m]}$ and $\hat{h}_{j, \boldsymbol{\sigma}}^{[m]}$. 
%Assume all base-learners are of the same form, i.e., $\hat{h}_{\boldsymbol{\mu}} = \hat{h}_{j, \boldsymbol{\mu}}$ and $\hat{h}_{\boldsymbol{\sigma}} = \hat{h}_{j, \boldsymbol{\sigma}}$ for all $j = 1, \cdots, J$, then
%\begin{align}
%\boldsymbol{u}_\mu^{[m]} &= \hat{h}_\mu^{[m]}(\boldsymbol{x}_{\cdot j}) + \boldsymbol{\epsilon}_\mu \\
%\boldsymbol{u}_\sigma^{[m]} &= \hat{h}_\sigma^{[m]}(\boldsymbol{x}_{\cdot j}) + \boldsymbol{\epsilon}_\sigma, \label{eq: u_si}
%\end{align}
%where $\boldsymbol{\epsilon}_\mu$ and $\boldsymbol{\epsilon}_\sigma$ are the residuals for both parameters respectively (for more details, see appendix \ref{apx: A}).
%Let $h_\mu(\boldsymbol{x}_{\cdot j})$ and $h_\sigma(\boldsymbol{x}_{\cdot j})$ for $j \in \{1, \cdots, J\}$ be the simple linear base-learners
The optimization process can then be divided into two parts: one is the ASL for the location parameter $\boldsymbol{\mu}$, and the other is for the scale parameter $\boldsymbol{\sigma}$.
As the ASL shrinks the optimal value, we consider only the optimal step-lengths for both parameters.

\subsubsection{Optimal step-length for $\boldsymbol{\mu}$}
The analytical optimal step-length for $\boldsymbol{\mu}$ in iteration $m$ is obtained through minimizing the empirical risk
\begin{align}
\nu^{*[m]}_{j^*, \boldsymbol{\mu}} &= \argmin_{\nu} \sum_{i=1}^n \rho\left(y_i, \{\hat{\eta}_{\boldsymbol{\mu}}^{[m]}(\boldsymbol{x}_{i \cdot}), \hat{\eta}_{\boldsymbol{\sigma}}^{[m-1]}(\boldsymbol{x}_{i \cdot})\} \right) \nonumber \label{opt: mu} \\
&= \argmin_{\nu} \sum_{i=1}^n \frac{\left(y_i - \hat{\eta}_{\boldsymbol{\mu}}^{[m-1]}(\boldsymbol{x}_{i \cdot}) - \nu \hat{h}_{j^*, \boldsymbol{\mu}}^{[m]}(x_{ij^*})\right)^2}{2\hat{\sigma}_i^{2[m-1]}},
\end{align}
where the expression $\hat{\sigma}_i^{2[m-1]}$ represents the square of the standard deviation in the previous iteration, i.e. $\hat{\sigma}_i^{2[m-1]} = (\hat{\sigma}_i^{[m-1]})^2$. 
The optimal value of $\nu_{j^*, \boldsymbol{\mu}}^{*[m]}$ is obtained by letting the derivative of the equation equal zero, so we get the analytical ASL for $\boldsymbol{\mu}$ (for more derivation details, see also appendix \ref{apx: A_mu}):
\begin{align}
\nu^{*[m]}_{j^*, \boldsymbol{\mu}} &= \frac{\sum_{i=1}^n \left(\hat{h}_{j^*, \boldsymbol{\mu}}^{[m]}(x_{ij^*})\right)^2}{\sum_{i=1}^n \frac{\left(\hat{h}_{j^*, \boldsymbol{\mu}}^{[m]}(x_{ij^*})\right)^2}{\hat{\sigma}_i^{2[m-1]}}} \label{asl: mu}.
\end{align}
It is obvious, that $\nu^{*[m]}_{j^*, \boldsymbol{\mu}}$ is not an independent parameter in GAMLSS but depends on the base-learner $\hat{h}_{\boldsymbol{\mu}}^{[m]}(x_{ij^*})$ with respect to the best performing variable $\boldsymbol{x}_{\cdot j^*}$ and the estimated variance in the previous iteration $\hat{\sigma}_i^{2[m-1]}$. 

In the special case of a Gaussian additive model, the scale parameter $\sigma$ is assumed to be constant, i.e. $\hat{\sigma}_i^{[m-1]} = \hat{\sigma}^{[m-1]}$ for all $i \in \{1, \cdots, n\}$. We then obtain
\begin{align}
\nu_{j^*, \boldsymbol{\mu}}^{*[m]} = \frac{\sum_{i=1}^n \left(\hat{h}_{j^*, \boldsymbol{\mu}}^{[m]}(x_{ij^*})\right)^2}{\frac{1}{\hat{\sigma}^{2[m-1]}}\sum_{i=1}^n \left(\hat{h}_{j^*, \boldsymbol{\mu}}^{[m]}(x_{ij^*})\right)^2} = \hat{\sigma}^{2[m-1]}. \label{asl: mu const}
\end{align}
This gives us an interesting property of the optimal step-length or ASL, i.e., the analytical ASL for $\mu$ in the Gaussian distribution is actually the variance (as computed in the previous boosting iteration).
This property enables this paper to be not only applicable for the special GAMLSS case, but also for the boosting of additive models with normal responses.
Therefore, in the case of Gaussian additive models, we can use $\nu_{j^*, \boldsymbol{\mu}}^{[m]} = \lambda \hat{\sigma}^{2[m-1]}$ as the step-length, which has a stronger theoretical foundation, instead of the common choice 0.1.

Back to the general GAMLSS case, we can further investigate the behavior of the step-length by considering the limiting case of $m \rightarrow \infty$. 
For large $m$, all  base-learner fits  $\hat{h}_{j^*, \boldsymbol{\mu}}^{[m]}(x_{ij^*})$ converge to zero or are similarly small. 
If we consequently approximate all $\hat{h}_{j^*, \boldsymbol{\mu}}^{[m]}(x_{ij^*})$ by some small  constant $h$, this gives an approximation of the analytical optimal step-length of
\begin{align}
\nu_{j^*, \boldsymbol{\mu}}^{*[m]} \approx \frac{\sum_{i=1}^n h^2}{\sum_{i=1}^n \frac{h^2}{\hat{\sigma}_i^{2[m-1]}}} = \frac{n h^2}{h^2 \sum_{i=1}^n \frac{1}{\hat{\sigma}_i^{2[m-1]}}} = \frac{n}{\sum_{i=1}^n \frac{1}{\hat{\sigma}_i^{2[m-1]}}}, \label{asl: mu harm}
\end{align}
which is  the harmonic mean of the estimated variances $\hat{\sigma}_i^{2[m-1]}$ in the previous iteration. 
While this expression requires  $m$ to be large, which may not be reached if  $m_{\text{stop}}$ is of moderate size to prevent overfitting, the expression still gives an indication of the strong dependence of the optimal step-length on the variances $\hat{\sigma}_i^{2[m-1]}$, which generalizes the optimal value of the additive model in (\ref{asl: mu const}).

\subsubsection{Optimal step-length for $\boldsymbol{\sigma}$} \label{sssec:asl_si}
%The other part of the optimization is to analyze the existence of the analytical ASL for the scale parameter $\sigma$ in GAMLSS with respect to Gaussian distribution and its properties. 

The optimal step-length for the scale parameter $\boldsymbol{\sigma}$ can be obtained analogously by minimizing the empirical risk, now with respect to $\nu_{j^*, \boldsymbol{\sigma}}^{*[m]}$. 
We obtain
\begin{align}
\nu_{j^*, \boldsymbol{\sigma}}^{*[m]} &= \argmin_{\nu} \sum_{i=1}^n \rho\left(y_i, \{\hat{\eta}_{\boldsymbol{\mu}}^{[m-1]}(\boldsymbol{x}_{i \cdot}), \hat{\eta}_{\boldsymbol{\sigma}}^{[m]}(\boldsymbol{x}_{i \cdot})\}\right) \nonumber \\
&= \argmin_{\nu} \sum_{i=1}^n \left(\hat{\eta}_{\boldsymbol{\sigma}}^{[m-1]}(\boldsymbol{x}_{i \cdot}) + \nu \hat{h}_{\boldsymbol{\sigma}}^{[m]}(x_{ij^*}) \right) + \sum_{i=1}^n \frac{\left(y_i - \hat{\eta}_{\boldsymbol{\mu}}^{[m-1]}(\boldsymbol{x}_{i \cdot})\right)^2}{2\exp\left(2\hat{\eta}_{\boldsymbol{\sigma}}^{[m-1]}(\boldsymbol{x}_{i \cdot}) + 2\nu \hat{h}_{\boldsymbol{\sigma}}^{[m]}(x_{ij^*}) \right)}. \label{opt: si}
\end{align}
After checking the positivity of the second-order derivative of the expression in equation \eqref{opt: si}, the optimal value can be obtained by setting the first-order derivative equal to zero:
\begin{align}
\sum_{i=1}^n \hat{h}_{\boldsymbol{\sigma}}^{[m]}(x_{ij^*}) - \sum_{i=1}^n \frac{\left(\hat{h}_{\boldsymbol{\sigma}}^{[m]}(x_{ij^*}) + \epsilon_{i, \boldsymbol{\sigma}} + 1 \right)\hat{h}_{\boldsymbol{\sigma}}^{[m]}(x_{ij^*})}{\exp\left(2\nu_{j^*, \boldsymbol{\sigma}}^{*[m]} \hat{h}_{\boldsymbol{\sigma}}^{[m]}(x_{ij^*}) \right)} \overset{!}{=} 0, \label{opt: si 0}
\end{align}
where $\epsilon_{i, \boldsymbol{\sigma}}$ denotes the residuals when regressing the negative partial derivatives $\boldsymbol{u}_{\boldsymbol{\sigma}, i}^{[m]}$ on the base-learner $\hat{h}_{\boldsymbol{\sigma}}^{[m]}(x_{ij^*})$, i.e.,$u_{\boldsymbol{\sigma}, i} = \hat{h}_{\boldsymbol{\sigma}}^{[m]}(\boldsymbol{x}_{i \cdot}) + \epsilon_{i, \boldsymbol{\sigma}}$.
Unfortunately, equation \eqref{opt: si 0} cannot be further simplified, which means that there is no analytical ASL for the scale parameter $\boldsymbol{\sigma}$ in the Gaussian distribution. 
Hence, the optimal ASL must be found by performing a conventional line search.
For more details, see also appendix \ref{apx: A_sigma}.

Even without an analytical solution, we can still use (\ref{opt: si 0}) to  further study the behavior of the ASL. Analogous to the derivation of (\ref{asl: mu harm}),  $\hat{h}_{\boldsymbol{\sigma}}^{[m]}(x_{ij^*})$ converges to zero for $m \rightarrow \infty$. If we approximate with a (small) constant  $\hat{h}_{\boldsymbol{\sigma}}^{[m]}(x_{ij^*}) \approx h, \forall i \in \{1, \cdots, n\}$. Then (\ref{opt: si 0}) simplifies to 
\begin{align}
& \sum_{i=1}^n h - \sum_{i=1}^n \frac{(h + \epsilon_{i, \boldsymbol{\sigma}} + 1)h}{\exp\left(2\nu_{j^*, \boldsymbol{\sigma}}^{*[m]} h\right)} = 0 \nonumber \\
%\Leftrightarrow & nh = \frac{\sum_{i=1}^n (h + 1 + \epsilon_{\sigma i}) h}{\exp(2\nu_\sigma^{[m]} h)} \\
%\Leftrightarrow & \exp(2\nu_\sigma^{[m]} h) = \frac{h \sum_{i=1}^n(h + 1 + \epsilon_{\sigma i})}{nh} \\
%\Leftrightarrow & \exp(2\nu_\sigma^{[m]} h) = h + 1 + \frac{1}{n}\sum_{i=1}^n \epsilon_{\sigma i} \\
\Leftrightarrow & \nu_{j^*, \boldsymbol{\sigma}}^{*[m]} = \frac{1}{2h} \log\left(h + 1 + \frac{1}{n}\sum_{i=1}^n \epsilon_{i, \boldsymbol{\sigma}}\right) \nonumber \\
\Leftrightarrow & \nu_{j^*, \boldsymbol{\sigma}}^{*[m]} = \frac{1}{2h} \log(h + 1), \label{opt: si log}
\end{align}
where $\frac{1}{n}\sum_{i=1}^n \epsilon_{i, \boldsymbol{\sigma}} = 0$ in the regression model. Equation (\ref{opt: si log}) can be further simplified by approximating the logarithm function with a Taylor series at $h = 0$, thus
\begin{align*}
\nu_{j^*, \boldsymbol{\sigma}}^{*[m]} &= \frac{1}{2h}\left(h - \frac{h^2}{2} + O(h^3) \right)\\
&= \frac{1}{2} - \frac{h}{4} + O(h^2).
\end{align*}
As $h \rightarrow 0$ for $m \rightarrow \infty$,  the limit of this approximate optimal step-length for $\sigma$ is %constant and equal to 0.5, i.e.,
\begin{align}
\lim_{m \rightarrow \infty} \nu_{j^*, \boldsymbol{\sigma}}^{*[m]} = \lim_{h \rightarrow 0} \frac{1}{2} - \frac{h}{4} = \frac{1}{2}. \label{asl: si 1/2}
\end{align}
Thus, the ASL for $\boldsymbol{\sigma}$ approaches approximately 0.05 if we take the shrinkage parameter $\lambda = 0.1$ and iterations run for a longer time (and the boosting algorithm is not stopped too early to prevent overfitting for this  trend to show).

\subsection{(Semi-)Analytical adaptive step-length}
Knowing the properties of the analytical ASL in boosting GAMLSS for the Gaussian distribution, we can replace the line search with the analytical solution for the location parameter $\boldsymbol{\mu}$. 
If we keep the line search for the scale parameter $\boldsymbol{\sigma}$, we call this the \textit{Semi-Analytical Adaptive Step-Length (SAASL)}. 
Moreover, we are interested in the performance of combining the analytical ASL for $\boldsymbol{\mu}$ with the approximate value $0.05 = \lambda \cdot \frac{1}{2}$ (with $\lambda = 0.1$) for the ASL for $\boldsymbol{\sigma}$, which is motivated by the limiting considerations discussed above and has a better theoretical foundation than selecting an arbitrary small value in the common FSL. 
We call this step-length setup  \textit{SAASL05}. 
In either of these cases, it is straightforward and computationally efficient to obtain the (approximate) optimal value(s) and both alternatives are faster than performing two line searches. 

The semi-analytical solution avoids the need for selecting a search interval for the line search, at least for the ASL for $\boldsymbol{\mu}$ in the case of SAASL and for both parameters for SAASL05. 
This is an advantage, since too large search intervals will cause additional computing time, but too small intervals may miss the optimal ASL value and again lead to an imbalance of updates between the parameters. 
Also note that the value 0.5 gives an indication for a reasonable range for the search interval for $\nu_{j^*, \boldsymbol{\sigma}}^{*[m]}$ if a line search is conducted after all . 

%The step length either found by doing a line search or calculated from the analytical expression (\ref{asl: mu}) should in fact be called more precisely the \textit{optimal step length}. This is a value that leads to the largest decrease possible of the empirical risk and usually leads to overfitting of the corresponding variable if no shrinkage is used [\cite{Hepp2016ApproachesTR}]. The actual adaptive step length we  apply in the boosting algorithm is  the product of two parts, the shrinkage parameter $\lambda$ and the optimal step length $\nu_\theta^*$. In the following, we refer by the term adaptive step length to 
%\begin{align*}
%\nu_\theta = \lambda \nu_\theta^*,
%\end{align*}
%and in this article, we take $\lambda = 0.1$. The adaptive step length is thus 10\% of the optimal step length. By comparison, the fixed step length $\nu = 0.1$ in Algorithm \ref{alg: ncgb} would correspond to a combination of shrinkage parameter $\lambda = 0.1$ with the optimal step length $\nu^*$ set to one.

The boosting GAMLSS algorithm with ASL for the Gaussian distribution is shown in Algorithm \ref{alg: saasl}.

\begin{breakablealgorithm}
\caption{Non-cyclical componentwise gradient boosting for the Gaussian distribution with different step-lengths - Extension of basic algorithm \ref{alg: ncgb}}
\label{alg: saasl}
\begin{algorithmic}[1]
\REQUIRE $\cdots$ Steps 1-7 equal to algorithm \ref{alg: ncgb} $\cdots$, in addition, choose shrinkage parameter $\lambda$.
\setalglineno{8}
\STATE Set or find the step-length $\nu_{j^*, \boldsymbol{\theta}_k}^{[m]}$ for $\boldsymbol{\theta}_k \in \{\boldsymbol{\mu}, \boldsymbol{\sigma}\}$ by one of the followings:
	\begin{itemize}
	\item Adaptive step-length (ASL):
		\begin{align*}
		\nu_{j^*, \boldsymbol{\theta}_k}^{*[m]} = \argmin_{\nu} \sum_{i=1}^n \rho \left(y_i, \hat{\eta}_{\boldsymbol{\theta}_k}^{[m-1]}(\boldsymbol{x}_{i \cdot}) + \nu \cdot \hat{h}_{j^*, \boldsymbol{\theta}_k}(x_{i j^*}) \right);
		\end{align*}
	\item Semi-analytical adaptive step-length (SAASL):
	
		if $\boldsymbol{\theta}_k = \boldsymbol{\mu}$, 
		\begin{align*}
		\nu_{j^*, \boldsymbol{\mu}}^{*[m]} = \frac{\sum_{i=1}^n \left(\hat{h}_{j^*, \boldsymbol{\mu}}(x_{i j^*})\right)^2}{\sum_{i=1}^n \frac{\left(\hat{h}_{j^*, \boldsymbol{\mu}}(x_{i j^*}\right)^2}{\hat{\sigma}_i^{2[m-1]}}},
		\end{align*}
		if $\boldsymbol{\theta}_k = \boldsymbol{\sigma}$, same as for ASL.
	\item Semi-analytical adaptive step-length (SAASL05):
		
		if $\boldsymbol{\theta}_k = \boldsymbol{\mu}$, same as for SAASL,
		
		if $\boldsymbol{\theta}_k = \boldsymbol{\sigma}$, $\nu_{j^*, \boldsymbol{\theta}_k}^{*[m]} = 0.5$.
		
	\end{itemize}
	and set adaptive step-length $\nu_{j^*, \boldsymbol{\theta}_k}^{[m]}$ as the optimal value with shrinkage $\lambda$:
		\begin{align*}
		\nu_{j^*, \boldsymbol{\theta}_k}^{[m]} = \lambda \cdot \nu_{j^*, \boldsymbol{\theta}_k}^{*[m]}.
		\end{align*}
\REQUIRE $\cdots$ Steps 9-13 equal to those in algorithm \ref{alg: ncgb} $\cdots$
\end{algorithmic}
\end{breakablealgorithm}

For a chosen shrinkage parameter of $\lambda = 0.1$, the $\nu_{\boldsymbol{\sigma}}$ in SAASL05 would be 0.05, which is a smaller or ``less aggressive" value than 0.1 in FSL, leading to a somewhat larger number of %That means the common step length setting 0.1 for the scale parameter $\sigma$ in Gaussian distribution is an aggressive one. Consequently, FSL theoretically shall need less 
boosting iterations but a smaller %until $\sigma$ is well estimated than SAASL05, but it increases the 
risk of overfitting, and to a better balance with the ASL for $\boldsymbol{\mu}$.
\section{Simulation Study}
\label{sec: sim}
In the following, two simulations are shown to demonstrate the performance of the adaptive algorithms. 
The first one compares the estimation accuracy between the different non-cyclical boosted GAMLSS algorithms with FSL or ASL in a Gaussian regression model for location and scale. 
The second one underlines the problem of FSL and the performance of the adaptive approaches  if the variance in this setting is large.

\subsection{Gaussian Location and Scale Model}
\label{sec: sim1}
The simulation study in \cite{Thomas2018} showed that their FSL non-cyclical approach outperforms the classical cyclical approach. 
We use the same setup to show that the ASL approach performs at least as good as the FSL non-cyclical approach (and hence also outperforms the classical cyclical approach). 
At the end of this subsection we will show that the reason for the good performance of FSL is due to the chosen simulated data structure.
The setup is the following: the response $y_i$ is drawn from $N(\mu_i, \sigma_i)$ for $n=500$ observations, with 6 informative covariates $x_{ij}, j \in \{1, \cdots, 6\}$ drawn independently from $\text{Uni}(-1, 1)$. 
The predictors of both distribution parameters are:
\begin{align*}
\eta_{\boldsymbol{\mu}}(\boldsymbol{x}_{i \cdot}) &= \mu_i = x_{i1} + 2 x_{i2} + 0.5 x_{i3} - x_{i4}\\
\eta_{\boldsymbol{\sigma}}(\boldsymbol{x}_{i \cdot}) &= \log(\sigma_i) = 0.5 x_{i3} + 0.25 x_{i4} - 0.25 x_{i5} - 0.5 x_{i6},
\end{align*}
where $x_3$ and $x_4$ are shared between both $\mu$ and $\sigma$. 
Moreover, $p_{\text{n-inf}} = 0, 50, 250 \text{ or } 500$ non-informative variables sampled from $\text{Uni}(-1, 1)$ are also added to the model. We conduct $B = 100$ simulation runs.

The estimated coefficients of $\eta_{\boldsymbol{\mu}}$ and $\eta_{\boldsymbol{\sigma}}$, whose values are taken at stopping iterations tuned by 10-fold CV with the maximum number of boosting iterations  set to 1000, are shown in Appendix Figures \ref{fig: sim1_muMat} and \ref{fig: sim1_siMat}.

Overall, the estimated coefficients are similar between all four methods, with the shrinkage bias of boosting only becoming apparent with an increasing number of noise variables. 

\begin{figure}[H]
\centering
\includegraphics[width=.8\textwidth]{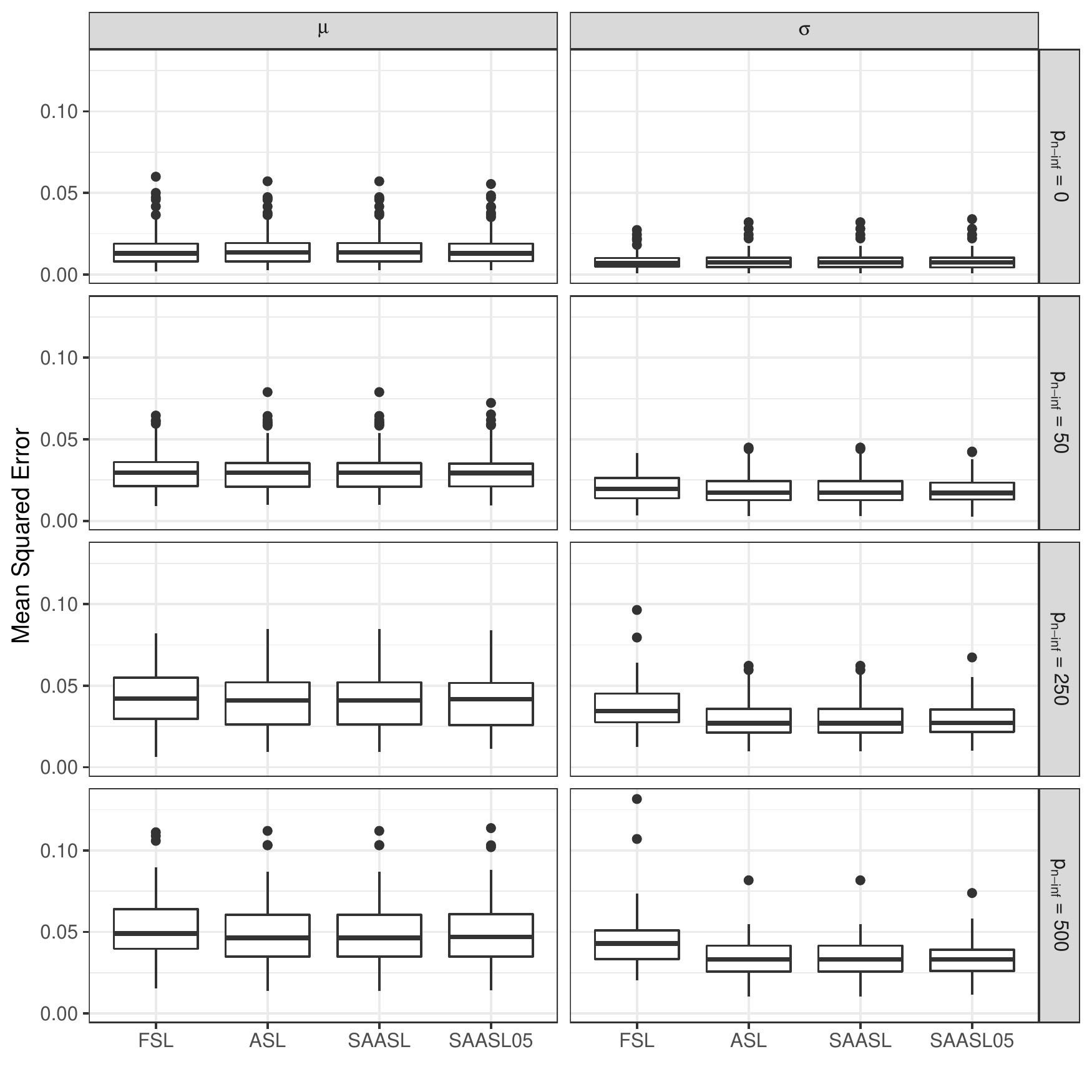}
\caption{Comparison between mean squared error for FSL and the three ASL methods. The left column comprises the MSE for $\eta_{\boldsymbol{\mu}}$, the right column for $\eta_{\boldsymbol{\sigma}}$. The different numbers of non-informative variables are represented row-wise.}
\label{fig: sim1_mse}
\end{figure}

Figure \ref{fig: sim1_mse} shows the comparison of the mean squared error (MSE) among non-cyclical boosted algorithms for $\boldsymbol{\mu}$ and $\boldsymbol{\sigma}$, where the MSEs are defined on the predictor level  as $\text{MSE}_{\boldsymbol{\mu}} = \frac{1}{n}\sum_{i=1}^n(\mu_i - \eta_{\boldsymbol{\mu}}(\boldsymbol{x}_{i \cdot}))^2$ and $\text{MSE}_{\boldsymbol{\sigma}} = \frac{1}{n}\sum_{i=1}^n(\log(\sigma_i) - \eta_{\boldsymbol{\sigma}}(\boldsymbol{x}_{i \cdot}))^2$, respectively. 
In general, all methods have a similar MSE, with the MSE of FSL increasing more strongly with the number of non-informative variables $p_{\text{n-inf}}$ and ASL methods hence slightly outperform FSL in the variance predictor for a high number of non-informative variables. 
ASL and SAASL show identical results, as they should if the line search is correctly conducted, with results returned by SAASL05 very similar.
%ASL and SAASL give identical results, as the optimal step length of $\mu$ lies in the default search interval $[0, 10]$ and the optimal step-length determined by line search in each iteration for ASL is thus the same as the analytical step-length in SAASL (cf. also Figure \ref{fig: sim1_stepLength}). 

\begin{figure}[h]
\centering
\includegraphics[width=.8\textwidth]{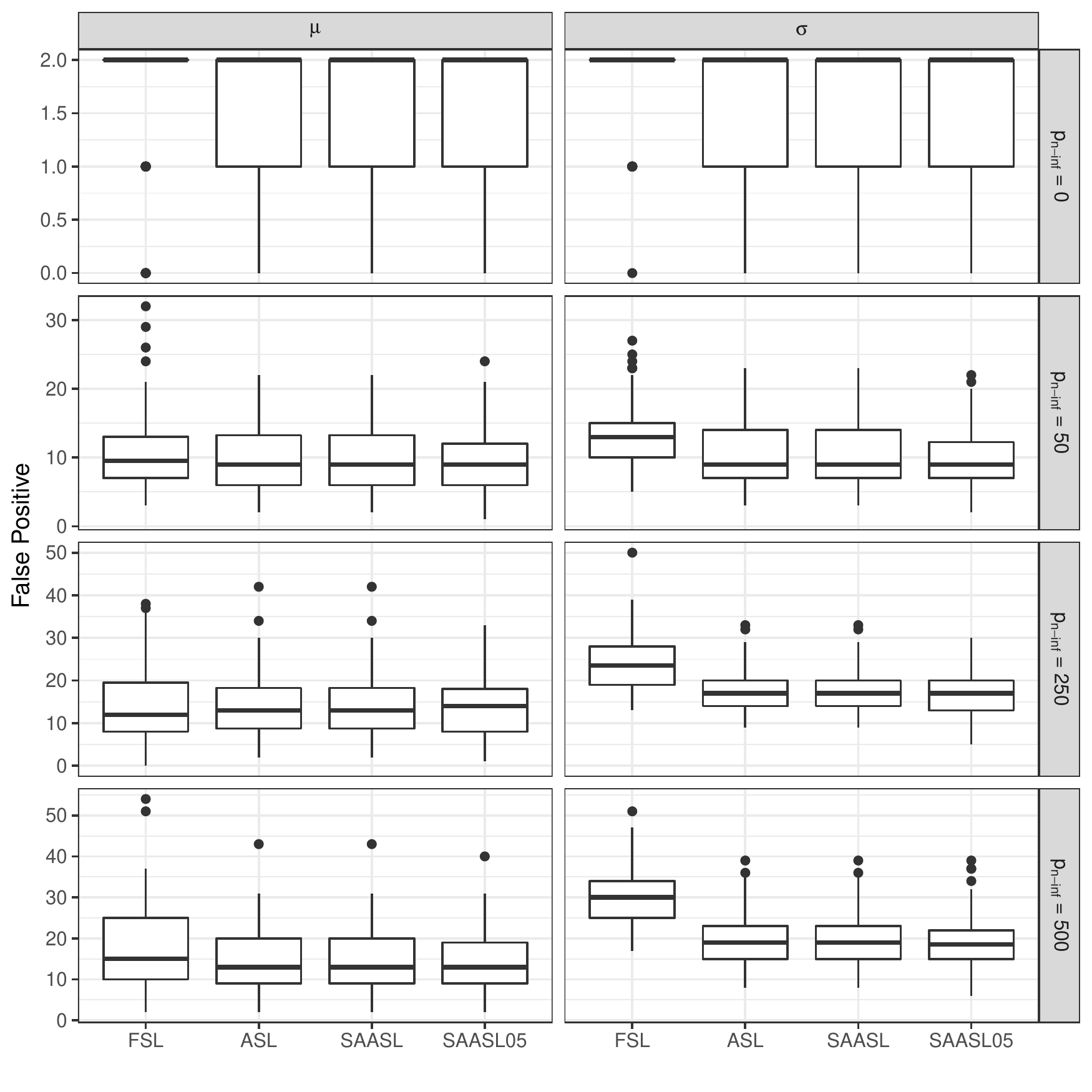}
\caption{Comparison between false positives for FSL and the three ASL methods. The left column comprises the false positives for $\boldsymbol{\mu}$, the right column for $\boldsymbol{\sigma}$. The different numbers of non-informative variables settings are represented row-wise.}
\label{fig: sim1_fp}
\end{figure}

Computing the negative log-likelihood in sample of the model fits reveals a slight advantage for FSL (see Appendix Figure \ref{fig: sim1_like}).
However, this can be linked to the fact that FSL selects more false positive variables on average than the adaptive approaches and thus shows a relatively stronger tendency to overfit the training data (Figure \ref{fig: sim1_fp}).
%The negative log-likelihood (see Appendix Figure \ref{fig: sim1_like}) is however smaller in-sample for FSL than for the adaptive methods, due to a larger number of false positives (overfitting, cf.\ also Figure  \ref{fig: sim1_fp}).

%Variable selection is one important reason to  use boosting as an estimation method, and the false positive rate is an important  measure for evaluation. Figure \ref{fig: sim1_fp} presents the number of false positives for each distribution parameter under different settings. For low-dimensional data, there is no great difference among methods for the false positives in $\mu$. When $p_{\text{n-info}}$ increases, however,
%The adaptive approaches select less false positive variables on average  than FSL. 
For $\boldsymbol{\sigma}$, even if $p_{\text{n-info}}$ is small, the false positive rates of the adaptive approaches are notably smaller than those of FSL. As discussed above,  $\nu_{j^*, \boldsymbol{\sigma}}^{[m]} \approx 0.05$ for large $m$ in the adaptive approach is smaller than $\nu_{\boldsymbol{\sigma}} = 0.1$ for FSL.
%, which means the predictor of $\sigma$ is updated in a more conservative manner than FSL. Each update of a predictor will influence the negative partial derivative and further affect the selection of other covariates. 
An update with a smaller, conservative step-length can apparently help to avoid overfitting and the adaptive step-length here seems to strike  the balance between learning speed and the number of false positives.
While it would also be possible to lower the step-length for FSL to reduce the number of 
%The low false positive rate makes the adaptive approaches very competitive over FSL. It is naturally possible to set an even smaller step length in FSL to prevent the 
non-informative variables included in the final model, this would increase the number of boosting iterations and the computing time, and it would not address the imbalance between updates for $\boldsymbol{\mu}$ and $\boldsymbol{\sigma}$.
The optimal choice of the step-length is also difficult without further tuning or an automatic selection as in ASL.

\begin{figure}[h]
\centering
\includegraphics[width=.7\textwidth]{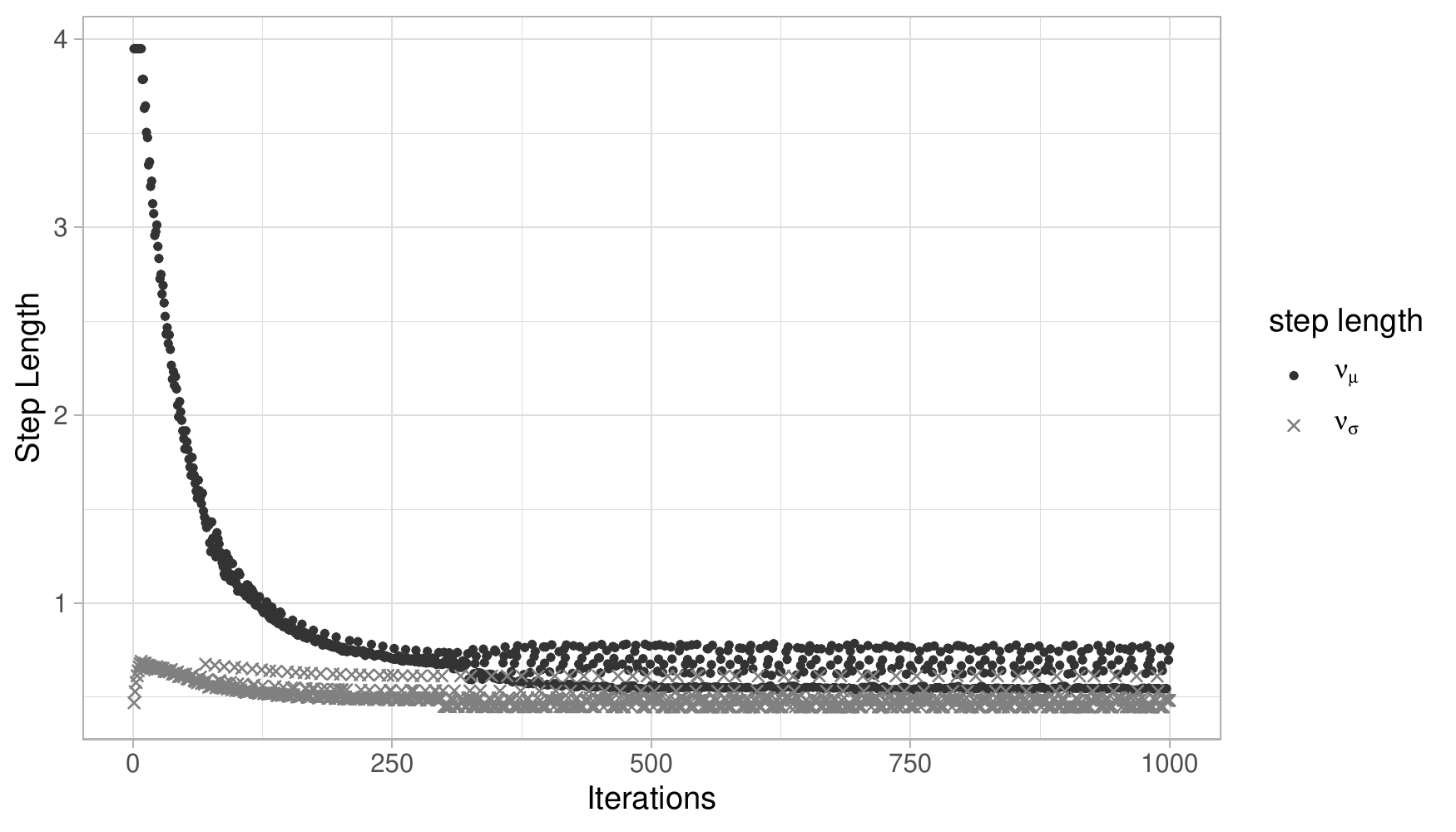}
\caption{Comparison of the optimal step-lengths $\nu_{j^*, \boldsymbol{\mu}}^{*[m]}$ and $\nu_{j^*, \boldsymbol{\sigma}}^{*[m]}$ in SAASL from one of the $100$ simulation runs. The step-lengths for $\mu$ are in black dots, the step-lengths for $\sigma$ in grey cross. Different horizontal layers of dots/crosses correspond to different covariates.}
\label{fig: sim1_stepLength}
\end{figure}

In Figure \ref{fig: sim1_stepLength} we show an example of the comparison between the optimal step-lengths in this case. 
As can be seen, the step-lengths for $\boldsymbol{\sigma}$ (depicted in grey) converge to $0.5$ as shown in section \ref{sssec:asl_si}. 
The second fact that becomes obvious when looking at the figure is that the optimal step-lengths for both predictors do not differ a lot. 
Even though differences can be observed in early iterations in particular, the step-lengths still have the same order of magnitude.
This is not only the case for this example but overall in this simulation setup. 
Having this in mind, the similar results for both approaches (FSL and ASL) are not very surprising anymore: there is hardly any difference in the approaches, since the updates do not need different step-lengths to be balanced. 
In the next subsection we will examine a case in which the data calls for different step-lengths, and see how both methods perform under those changed circumstances. 

\subsection{Large Variance with resulting Imbalance between Location and Scale}
\label{sec: sim2}

As discussed above, the Gaussian location and scale model in section \ref{sec: sim1} did not lead to a large difference between FSL and ASL, as the optimal step-lengths for $\boldsymbol{\mu}$ and $\boldsymbol{\sigma}$ were roughly similar and the imbalance between the updates for the two predictors in FSL was thus not large. 
In this section, we investigate a setting with a large variance, which leads to a stronger imbalance between the two parts of the model. 

In the following, we use SAASL as a representative of the adaptive approaches in our presentation, as it yields identical results to ASL, but avoids the numerical search for the optimal $\nu_{\boldsymbol{\mu}}$ by using the analytical result \eqref{asl: mu}.
Since estimated effects generally deviated more strongly from the theoretical values than before due to the large variance (details will be discussed later), 
we additionally compared the results to those obtained using GAMLSS with penalized maximum likelihood estimation as implemented in the R-package \texttt{gamlss}  [\cite{doi:10.1111/j.1467-9876.2005.00510.x}].

Consider the data generating mechanism $y_i \sim N(\mu_i, \sigma_i), i \in \{1, \cdots, 500\}$ with $B = 100$ simulation runs. The predictors are determined by
\begin{align*}
\eta_{\boldsymbol{\mu}}(\boldsymbol{x}_{i \cdot}) &= \mu_i = 1 + x_{i1} + 2x_{i2} - x_{i3}\\
\eta_{\boldsymbol{\sigma}}(\boldsymbol{x}_{i \cdot}) &= \log(\sigma_i) = 5 + 0.1x_{i1} - 0.2x_{i2} + 0.1x_{i3},
\end{align*}
where $\boldsymbol{x}_{\cdot j} \sim \text{Uni}(-1, 1), j \in \{1, 2, 3, 4, 5\}$, $\boldsymbol{x}_{\cdot 4}$ and $\boldsymbol{x}_{\cdot 5}$ are noise variables. 
Note that this choice of $\eta_{\boldsymbol{\sigma}}$ leads to an extremely large standard deviation in the order of 150 due to the large intercept 5.
The stopping iteration is obtained by 10-fold CV, and the maximum number of iterations is 3000 and 2,000,000 for SAASL and FSL respectively.
%Except for one simulation run, the stopping iterations of others determined by CV for FSL lie within two millions iterations.
%Coefficients for $\mu$ and $\sigma$ are taken at the corresponding stopping iteration of each simulation run.

\begin{figure}[h]
\centering
\includegraphics[width=\textwidth]{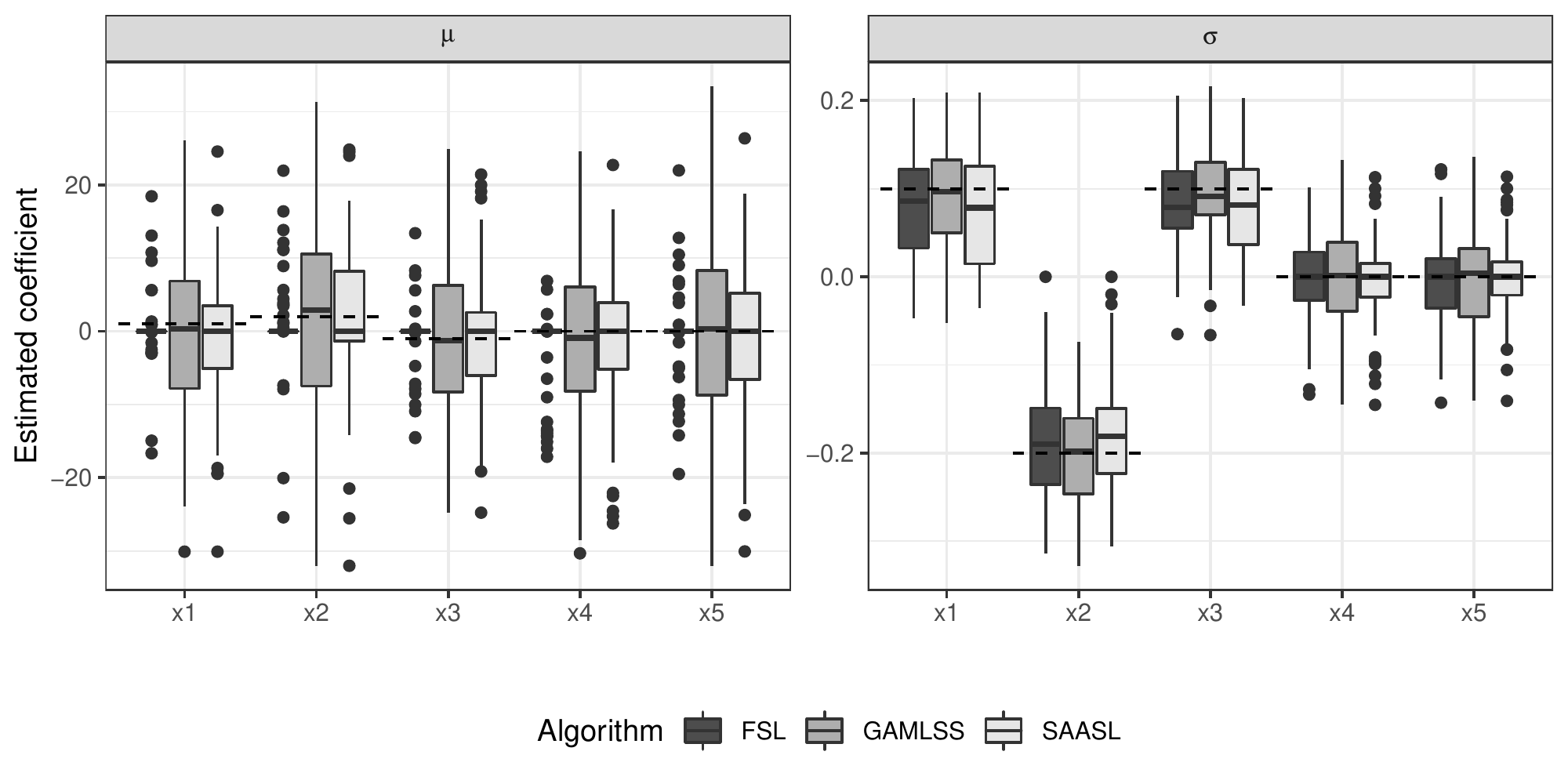}
\caption{Distribution of coefficient estimates from $B = 100$ simulation runs. The true coefficients are marked by the dashed horizontal lines.}
\label{fig: box_fgs}
\end{figure}

As can be seen in Figure \ref{fig: box_fgs}, both fixed and adaptive step-lengths yield reasonable estimates regarding $\eta_{\boldsymbol{\sigma}}$, but FSL results in many false negative estimates equal to zero for $\eta_{\boldsymbol{\mu}}$ in the majority of the simulation runs.
This is of course connected to the relative importance of the variance component in this setting, which should in itself already lead to a preference for updating $\eta_{\boldsymbol{\sigma}}$ rather than $\eta_{\boldsymbol{\mu}}$ in early boosting iterations due to the fact that the negative gradient for $\boldsymbol{\mu}$ (i.e. $u_{\boldsymbol{\mu}, i} = \sum_{i=1}^n (y_i - \mu_i)/\sigma_i^2 \text{ with large } \sigma_i$) is actually scaled by the variance (recall the large intercept 5, log-link and the resulting exponential transformation) and hence very small.
As a consequence, the impact on the global loss of base-learners fit to the gradient is also small compared to those suggested for updates regarding $\sigma$ in step \ref{alg: ncgb10} of Algorithm \ref{alg: ncgb}.
Then, using the same step-length for both parameters makes it clearly harder to identify informative effects on $\mu$ as they are trivialized in comparisons.

\begin{figure}[h]
\centering
\includegraphics[width=\textwidth]{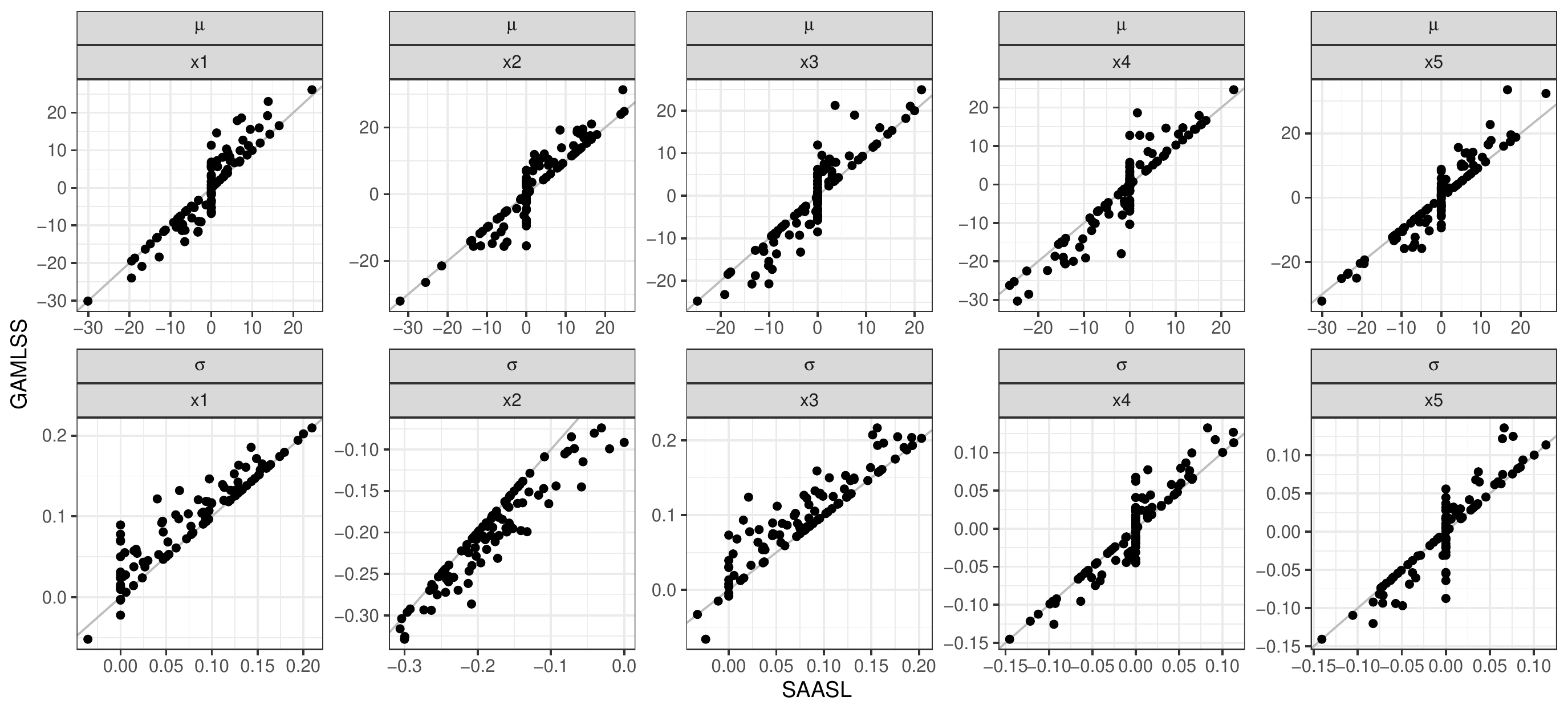}
\caption{Pairwise comparison of the estimated coefficients between GAMLSS and SAASL.}
\label{fig: svg_mu}
\end{figure}

The adaptive step-lengths implemented in SAASL compensates for this disadvantage.
Compared to the simulation results in the previous subsection the estimates regarding $\eta_{\boldsymbol{\mu}}$ are less precise with large variability around the true values. This is not a problem of SAASL but again the consequence of the large variance, obscuring the effects on the mean, and also encountered using the penalized maximum likelihood approach implemented in the \texttt{gamlss}-package (called GAMLSS in Figure \ref{fig: box_fgs}).
%Overall, the estimation of $\eta_\mu$ in FSL is not acceptable. SAASL also gives less precise estimates in this difficult setting than before, with a large variance around the true values. 
%However, this is only a consequence of the large variance, which makes it difficult to identify the relatively small effects on the expected value.
%This is also evident using the penalized maximum likelihood approach implemented in the \texttt{gamlss}-package (Figure \ref{fig: box_fgs}).
%However, results are comparable to the ones from GAMLSS based on penalized maximum likelihood estimation, i.e.\ the large variability in the estimates is due to the large variability in the responses. 
The variability in the estimates is actually somewhat smaller than for GAMLSS due to the regularization inherent in the boosting approach. 
This is also illustrated in Figure \ref{fig: svg_mu} in the pairwise comparison of the estimated coefficients for both methods, where SAASL leads to similar but slightly closer to zero estimates compared to the penalized maximum likelihood based method GAMLSS.

Interestingly, Figure \ref{fig: box_fgs} also reveals that the inability to identify the informative variables results in the lowest MSE for all three individual coefficients for $\boldsymbol{\mu}$ when using FSL (for more numerical details, see Appendix \ref{apx: mse_table}).
As can be seen from Table \ref{tbl: sim2_mspe}, however, the combined additive predictor performs worse in terms of overall MSE than both GAMLSS and SAASL, with the latter performing best.

%But a problem from Figure \ref{fig: box_fgs} arises that FSL actually has the smaller MSE for the coefficients of $\boldsymbol{\mu}$. 
%Table \ref{tbl: sim2_coef_mse} lists the details of the average MSE for all simulations and the MSE of the common simulations, in which the corresponding covariable is selected. 
%There is not much difference in the comparison for $\boldsymbol{\sigma}$. 
%But in either the total simulations or the true positive subset, FSL has the smallest average MSE for all three coefficients for $\boldsymbol{\mu}$. 
%The performance of SAASL is bette than GAMLSS.
%But as mentioned above, FSL is doing better here for the ``wrong" reasons, i.e., it has a poorer variable selection behavior (we'll discuss it later). 
%Moreover, smaller MSE for each coefficient of $\boldsymbol{\mu}$ in FSL does not guarantee a smaller in-sample MSE for the responses.
%Table \ref{tbl: sim2_mspe} shows the summary of the in-sample MSE for each approach. 
%From this table we can find that even the SAASL has a larger variance in the estimated coefficients (Figure \ref{fig: box_fgs}) and the performance of the average MSE for each coefficient is worse than FSL (Table \ref{tbl: sim2_coef_mse}), its in-sample MSE performance is still the best among the three approaches.

\begin{table}[h]
\centering
\caption{Summary of the in-sample MSE for each estimation methods, i.e. $\frac{1}{n}\sum_{i=1}^n\left(y_i - \hat{y}_i\right)^2$.} \label{tbl: sim2_mspe}
\begin{tabular}{c|cccccc}
  \toprule
 & Min. & 1st Qu. & Median & Mean & 3rd Qu. & Max. \\ 
  \midrule
     FSL & 19848   & 21796   & 22547   & 22688   & 23579   & 27026   \\ 
      GAMLSS & 19707   & 21687   & 22414   & 22586   & 23515   & 26883   \\ 
      SAASL & 19679   & 21663   & 22372   & 22554   & 23443   & 26883   \\ 
   \bottomrule
\end{tabular}
\end{table}

\begin{figure}[h]
\centering
\begin{subfigure}[b]{.48\linewidth}
\includegraphics[width=\textwidth]{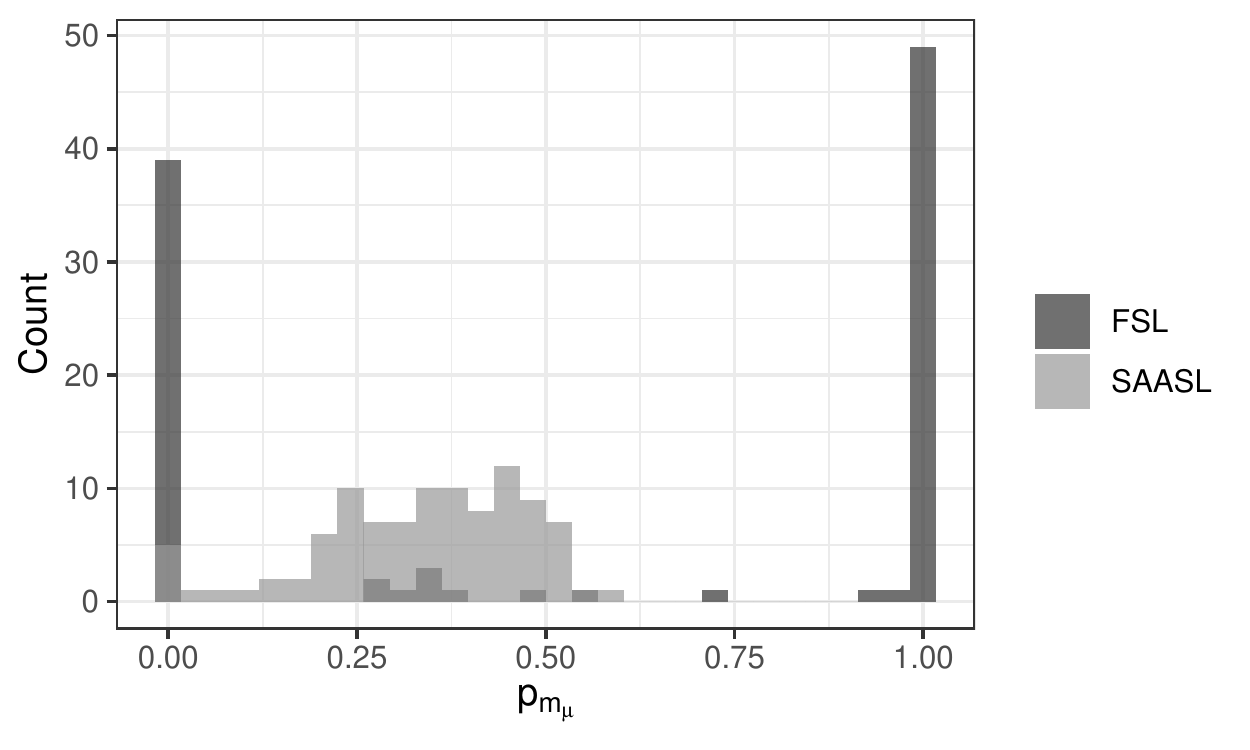}
\caption{Histogram} \label{fig: sim2_withNoise_muvsiHist}
\end{subfigure}
\hfill
\begin{subfigure}[b]{.48\linewidth}
\includegraphics[width=\textwidth]{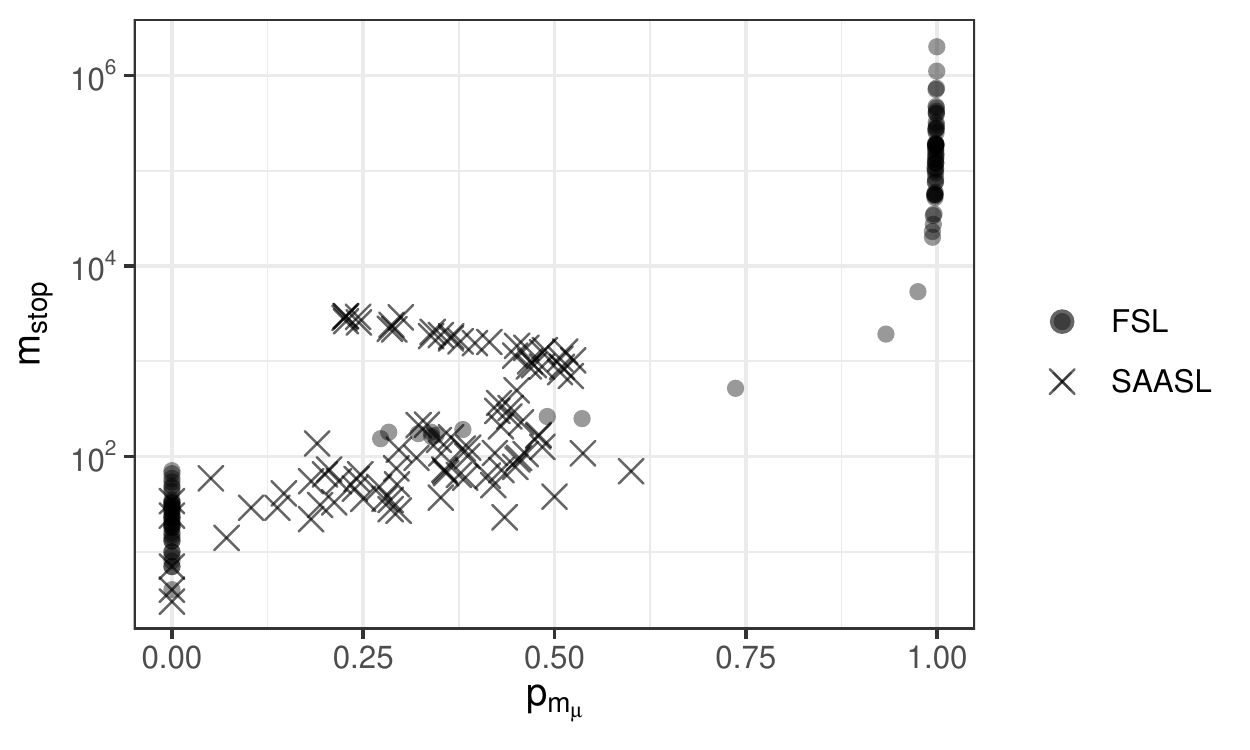}
\caption{Scatter plot} \label{fig: sim2_withNoise_muvsiPoint}
\end{subfigure}
\caption{Distribution of $p_{m_{\boldsymbol{\mu}}}$ in $B = 100$ simulation runs. (\subref{fig: sim2_withNoise_muvsiHist}) Histogram of $p_{m_{\boldsymbol{\mu}}}$. The histogram of the two approaches are overlayed using transparency. (\subref{fig: sim2_withNoise_muvsiPoint}) Scatter plot of $m_{\text{stop}}$ against $p_{m_{\boldsymbol{\mu}}}$. Points and crosses are displayed with transparency. The $y$-axis is displayed on a logarithmic scale with base 10. Each tick represents a tenfold increase over the previous one.} \label{fig: sim2_withNoise_muvsi}
\end{figure}

To further highlight the differences in the selection behavior between FSL and SAASL, Figure \ref{fig: sim2_withNoise_muvsi} illustrates the proportion of boosting iterations used to update $\boldsymbol{\mu}$ over the course of the model fits, i.e. $p_{m_{\boldsymbol{\mu}}} = m_{\boldsymbol{\mu}} / (m_{\boldsymbol{\mu}} + m_{\boldsymbol{\sigma}})$, where $m_{\boldsymbol{\mu}} + m_{\boldsymbol{\sigma}} = m_{\text{stop}}$.
The bimodal distribution for FSL observed in the histogram in panel (\subref{fig: sim2_withNoise_muvsiHist}) demonstrates another problem of the fixed step-lengths in this setting.
Considering the many estimates equal or close to zero observed in Figure \ref{fig: box_fgs}, the mode close to $p_{m_{\boldsymbol{\mu}}} = 0$ is expected, as it describes the proportion of simulation runs where $\mu$ has not been updated at all.
However, as soon as at least one base-learner for $\boldsymbol{\mu}$ is recognized as an effective model parameter, the small step-length fixed at 0.1 requires a huge number of updates for the base-learner to actually make an impact on the global loss (hence the large number of maximum iterations allowed for FSL).
This results in the second mode also around $p_{m_{\boldsymbol{\mu}}} = 1$, as the algorithm is mainly occupied with $\mu$ in the corresponding runs.

%Figure \ref{fig: sim2_withNoise_muvsi}(\subref{fig: sim2_withNoise_muvsiHist}) presents the histogram of the proportion of the number of boosting iterations used for updating $\mu$ up to $m_{\text{stop}}$ for all 100 simulation runs, i.e. $p_{m_{\mu}} = m_\mu / (m_\mu + m_\sigma)$, where $m_\mu + m_\sigma = m_{\text{stop}}$.
%The analogical histogram for $p_{m_\sigma}$ can be omitted, as $p_{m_\mu} + p_{m_\sigma} = 1$ and it is only a plot whose axis label with 0 and 1 reversed.
%A bimodal distribution for FSL can be observed with many values closed to 0 and a second mode closed to 1.
%The mode closed to 0 indicates either the predictor for $\mu$ in lots of simulation runs was not updated at all, or the boosting procedure stopped after only few updates for $\mu$ (which actually does not appear in this example).
%No matter in which case, $\mu$ is considered as a noise parameter by FSL, and therefore stopped after acquiring enough knowledge about $\sigma$. 
%As long as $\mu$ is recognized as a effective model parameter, FSL goes to the other extreme $p_{m_{\mu}}$ closed to 1, i.e., most of boosting iterations are used to update $\mu$. 
%Because the update of $\eta_\mu$ in each iteration with small fixed step-length setting is very small, hundred thousands or even millions iterations are required to get relatively well fitted coefficients of $\mu$ in this example (cf. also figure \ref{fig: sim2_withNoise_muvsi}(\subref{fig: sim2_withNoise_muvsiPoint})).

This is illustrated by the scatter plot in Figure \ref{fig: sim2_withNoise_muvsi}(\subref{fig: sim2_withNoise_muvsiPoint}), where $p_{m_{\boldsymbol{\mu}}}$ is plotted against the stopping iteration $m_{\text{stop}}$.
%Figure \ref{fig: sim2_withNoise_muvsi}(\subref{fig: sim2_withNoise_muvsiPoint}) presents a scatter plot of $p_{m_\mu}$ against the stopping iteration $m_{\text{stop}}$.
Note that the y-axis is displayed with a logarithmic scale and each tick on the y-axis represents a tenfold increase over the previous one.
The few points (FSL), whose $m_{\text{stop}}$ lie between $10^2$ and $10^3$, show a better balance between the updates of $\boldsymbol{\mu}$ and $\boldsymbol{\sigma}$ than other points, i.e., the middle region of $p_{m_{\boldsymbol{\mu}}}$.
But we also observe a bimodal distribution for FSL, i.e., lots of points are equal or close to $p_{m_\mu} = 0$ and 1, with very low and extremely large values for $m_{\text{stop}}$ resulting, respectively.
%But as mentioned above, such number of iterations are far less enough to extract sufficient information about $\mu$ from data.
%Thus from the practical point of view these points make no difference from the points closed to $p_{m_\mu} = 0$.

For SAASL, the distribution of $p_{m_{\boldsymbol{\mu}}}$ in Figure \ref{fig: sim2_withNoise_muvsi}(\subref{fig: sim2_withNoise_muvsiHist}) is unimodal.
The mode smaller than 0.5 indicates SAASL updates $\boldsymbol{\sigma}$ a little more frequently than $\boldsymbol{\mu}$.
Unlike the cyclical approach that enforces an equal number of updates for all distribution parameters, the balance formed by SAASL is more natural.
This balance enables the alternate updates between two predictors even though they lie on different scales.
Therefore, the information in $\boldsymbol{\mu}$ can be fairly discovered in time and it reduces the risk of overlooking the informative base-learners with respect to $\boldsymbol{\mu}$.
The number of simulations runs, in which $\boldsymbol{\mu}$ is not updated at all ($p_{m_{\boldsymbol{\mu}}} = 0$), reduces from 39 in FSL to only 5 in SAASL. 
Moreover, none of the 100 simulations require a substantial amount of updates for $\boldsymbol{\mu}$ to get well estimated coefficients (cf. also Figure \ref{fig: box_fgs}).

Table \ref{tbl: sim2_fp} displays the information about false positives and false negatives of the two approaches in all 100 simulations with respect to $\boldsymbol{\mu}$ and $\boldsymbol{\sigma}$.
For example, the second and fourth number 77 and 21 in the first line indicate that the informative variable $\boldsymbol{x}_{\cdot 2}$ is not included in the final model in 77 out of 100 simulation runs (i.e. false negative), while there are 21 simulations whose final model contains the non-informative variable $\boldsymbol{x}_{\cdot 4}$ (i.e. false positive).
Similar as Figure \ref{fig: sim1_fp} in section \ref{sec: sim1}, the conservative small step-length for $\boldsymbol{\mu}$ in FSL increases the number of boosting iterations, but reduces the risk of overfitting. 
Less simulations containing the noise variables for $\boldsymbol{\mu}$ in FSL than in SAASL confirms this behavior.
According to equation \eqref{asl: si 1/2} the ASLs $\nu_{j^*, \boldsymbol{\sigma}}$ are a sequence of values around 0.05, and (except for the values at early boosting iterations) most of them smaller than 0.1. 
There are correspondingly slightly more simulations in FSL overfitting the $\boldsymbol{\sigma}$-submodel than in SAASL.

\begin{table}[h]
\centering
\caption{The number of simulations with false positives and false negatives for each variable under different modelling methods with respect to the two model parameters. The false negatives part shows the number of simulations in which the informative variables are excluded from the final model, and the false positives part shows how many simulations include the non-informative variables in their final model. Values are taken at the stopping iteration determined by 10-fold CV.} \label{tbl: sim2_fp}
\begin{tabular}{cc|ccc|cc}
\toprule
	&	&	\multicolumn{3}{|c|}{False Negatives}	&	\multicolumn{2}{|c}{False Positives}\\
	&	&	$\boldsymbol{x}_{\cdot 1}$	&	$\boldsymbol{x}_{\cdot 2}$	&	$\boldsymbol{x}_{\cdot 3}$	&	$\boldsymbol{x}_{\cdot 4}$	&	$\boldsymbol{x}_{\cdot 5}$ \\
\midrule
\multirow{2}{*}{$\mu$} & FSL	&	83	&	77	&	81	&	21	&	20	\\
& SAASL	&	28	&	24	&	28	&	72	&	73	\\
\hline
\multirow{2}{*}{$\sigma$} & FSL	&	9	&	1	&	6	&	83	&	82	\\
& SAASL	&	18	&	1	&	9	&	70	&	67	\\
\bottomrule
\end{tabular}
\end{table}

Although non-informative variables of $\boldsymbol{\mu}$ are excluded from the FSL model, the informative ones are excluded as well.
Actually $\boldsymbol{\mu}$ was not updated in many simulations at all (cf. Figure \ref{fig: sim2_withNoise_muvsi}(\subref{fig: sim2_withNoise_muvsiHist})).
The false negatives part of Table \ref{tbl: sim2_fp} for $\boldsymbol{\mu}$ confirms this.
The informative variables $\boldsymbol{x}_{\cdot 1}$ to $\boldsymbol{x}_{\cdot 3}$ are excluded from the final model in the majority of simulations with FSL but not with SAASL.
For $\boldsymbol{\sigma}$, the smaller step-length $\nu_{j^*, \boldsymbol{\sigma}}$ in SAASL selects variables more conservatively and as a consequence slightly more simulations underfit the $\boldsymbol{\sigma}$-submodel in SAASL than in FSL, but the difference is far less pronounced.

\section{Applications}
\label{sec: empirical}
We apply the proposed algorithms  to two datasets. 
The malnutrition dataset demonstrates the shortcomings of FSL and the pitfalls of using numerical determination of ASL with a fixed search interval, and with the riboflavin dataset we illustrate the variable selection properties of each algorithm.
\subsection{Malnutrition of children in India}
The first data called \texttt{india} from the R package \texttt{gamboostLSS} [\cite{rgamlss, india}] are sampled from the Standard Demographic and Health Survey between 1998 and 1999 on malnutrition of children in India [\cite{india}]. 
The sample contains 4000 observations and four variables (BMI of the child (cBMI), age of the child in months (cAge), BMI of the mother (mBMI) and age of the mother in years (mAge)). 
The outcome of interest in this case is a numeric z-score for malnutrition ranging from -6 to 6, where the negative values represent malnourished children. 
To highlight the problems of using a fixed step-length, we work with the original variable stunting (corresponding to 100 * z-score). 
The identity and logarithm functions are used as the link functions for $\boldsymbol{\mu}$ and $\boldsymbol{\sigma}$ respectively.

Because this is not a high-dimensional data example, we use the GAMLSS with penalized maximum-likelihood estimation as a gold standard to examine the effectiveness of the adaptive approaches.

Table \ref{tbl: india} lists the estimated coefficients of each variable on the predictors $\eta_{\boldsymbol{\mu}}$ and $\eta_{\boldsymbol{\sigma}}$ at the stopping iteration tuned by 10-folds CV, where the maximum number of iterations is set to 2000. 
The estimated intercept in $\eta_{\boldsymbol{\sigma}}$ indicates a large variance of the response, with the setting thus being similar to the second simulation above.
It is therefore not surprising that FSL selects only one variable (cAge) for $\eta_{\boldsymbol{\mu}}$, i.e. a large number of updates for the base-learner are required but the given maximal boosting iteration is not large enough.
In practice we can certainly increase the maximum number of iterations as well as enlarge the commonly applied step-length 0.1 in order to estimate the coefficients well.
But their choices are very subjective and probably result in tedious manual fine-tuning based on trial and error.

The ASL method with the default predefined search interval $[0, 10]$ encounters a similar problem as FSL.
Apart from the only selected and underfitted variable cAge for $\boldsymbol{\mu}$, the two variables (cBMI and cAge) for the $\boldsymbol{\sigma}$-submodel are also underfitted compared with the results from the gold standard GAMLSS.
The reason for this phenomenon lies in the relationship between the variance and step-length discussed in equation \eqref{asl: mu}.
The log-link or exponential transformation for $\eta_{\boldsymbol{\sigma}}$ in this example data requires a sequence of huge step-lengths, but the default search interval does not fulfill this requirement.

An estimation of ASL by increasing its search interval to $[0, 50000]$, denoted as ASL5 in Table \ref{tbl: india}, results in coefficients comparable to those of GAMLSS.
But choosing a suitable search interval becomes an unavoidable side task for ASL when analyzing this kind of dataset.

The results of the two semi-analytical approaches hardly differ from the maximum likelihood based GAMLSS. Unlike the numerical determination with a fixed search interval in ASL, the analytical approaches replace this procedure with a direct and precise solution that gets rid of the potential manual intervention (e.g. increasing the search interval). 
Contrary to the direct influence of the variance on $\nu_{j^*, \boldsymbol{\mu}}^{*[m]}$ in equation \eqref{asl: mu}, the optimal step-length $\nu_{j^*, \boldsymbol{\sigma}}^{*[m]}$ is dominated by the chosen base-learner, but as the number of learning iterations increases, such effects gradually disappear, and $\nu_{j^*, \boldsymbol{\sigma}}^{*[m]}$ finally converges to 0.5.
Thus, our default search interval $[0, 1]$ is sufficient for $\nu_{j^*, \boldsymbol{\sigma}}^{*[m]}$, and increasing the range of search interval as for $\nu_{j^*, \boldsymbol{\mu}}^{*[m]}$ in ASL is almost never necessary.

Theoretically, the ASL with a sufficiently large search interval (ASL5 in this example) and SAASL should result in the same values as discussed in the previous theoretical section.
Due to the calculation accuracy of computers and the numerical optimization steps, their outputs are very similar but can differ slightly for the malnutrition data.

%Moreover, because the default predefined searching interval ($[0, 10]$) is not large enough, ASL also selects only the age of child variable for $\eta_\mu$. Additionally, both FSL and ASL underestimate the coefficient of cAge  compared with the semi-analytical solutions and the results from GAMLSS.

\begin{table}[ht]
\caption{Comparison of the estimated coefficients.}
\label{tbl: india}
\centering
\begin{tabular}{rlrrrrrr}
  \toprule
 &  & FSL & ASL & ASL5 & SAASL & SAASL05 & GAMLSS \\ 
  \midrule
  \hline
\multirow{2}{*}{(Intercept)} & $\eta_{\boldsymbol{\mu}}$ & -174.77179 & -169.20269 & -91.16032 & -91.16011 & -91.16041 & -91.15953 \\ 
 & $\eta_{\boldsymbol{\sigma}}$ & 4.88080 & 4.87412 & 4.91206 & 4.91207 & 4.91206 & 4.91205 \\ 
 \hline
  \multirow{2}{*}{cBMI} & $\eta_{\boldsymbol{\mu}}$ & 0.00000 & 0.00000 & -13.92524 & -13.92509 & -13.92526 & -13.92552 \\ 
 & $\eta_{\boldsymbol{\sigma}}$ & -0.00319 & -0.00268 & -0.01527 & -0.01526 & -0.01527 & -0.01526 \\ 
 \hline
  \multirow{2}{*}{cAge} & $\eta_{\boldsymbol{\mu}}$ & -0.03845 & -0.37092 & -5.84669 & -5.84665 & -5.84670 & -5.84663 \\ 
 & $\eta_{\boldsymbol{\sigma}}$ & -0.00103 & -0.00074 & 0.00284 & 0.00284 & 0.00284 & 0.00284 \\ 
 \hline
  \multirow{2}{*}{mBMI} & $\eta_{\boldsymbol{\mu}}$ & 0.00000 & 0.00000 & 11.70780 & 11.70770 & 11.70782 & 11.70787 \\ 
 & $\eta_{\boldsymbol{\sigma}}$ & 0.00897 & 0.00875 & 0.00864 & 0.00864 & 0.00864 & 0.00864 \\ 
 \hline
  \multirow{2}{*}{mAge} & $\eta_{\boldsymbol{\mu}}$ & 0.00000 & 0.00000 & 0.02570 & 0.02564 & 0.02571 & 0.02575 \\ 
 & $\eta_{\boldsymbol{\sigma}}$ & 0.00537 & 0.00513 & 0.00530 & 0.00530 & 0.00530 & 0.00530 \\ 
 \hline
   \bottomrule
\end{tabular}
\end{table}

\begin{figure}[h]
\centering
\begin{subfigure}[b]{.48\linewidth}
\includegraphics[width=\textwidth]{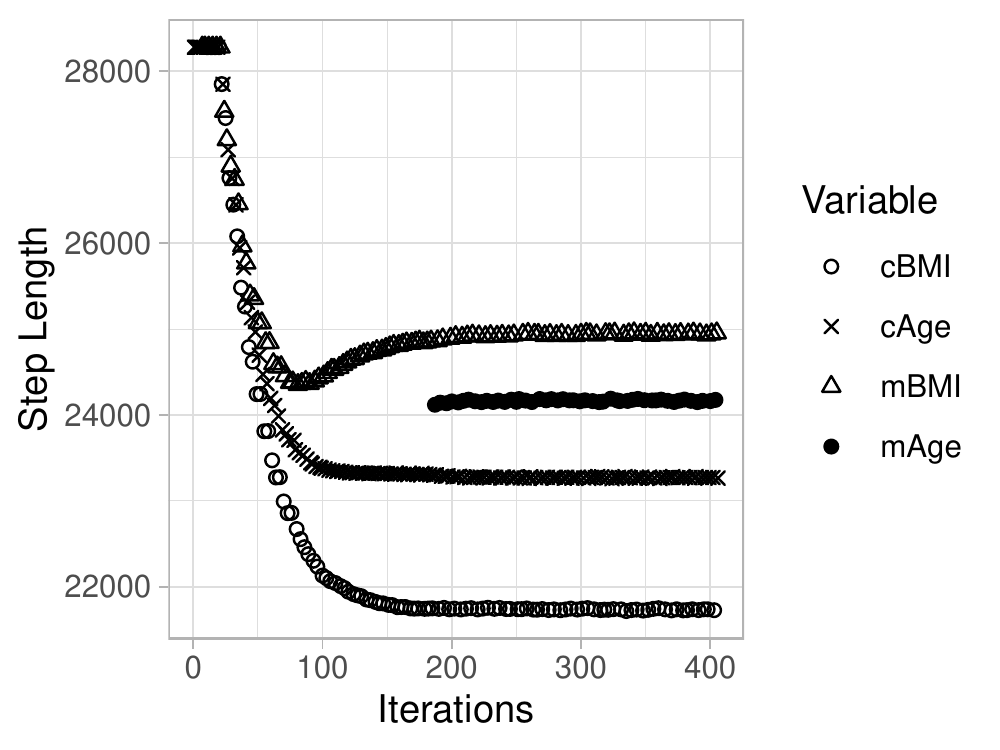}
\caption{$\nu_{j^*, \boldsymbol{\mu}}^{*[m]}$} \label{fig: india_saasl_v_mu}
\end{subfigure}
\begin{subfigure}[b]{.48\linewidth}
\includegraphics[width=\textwidth]{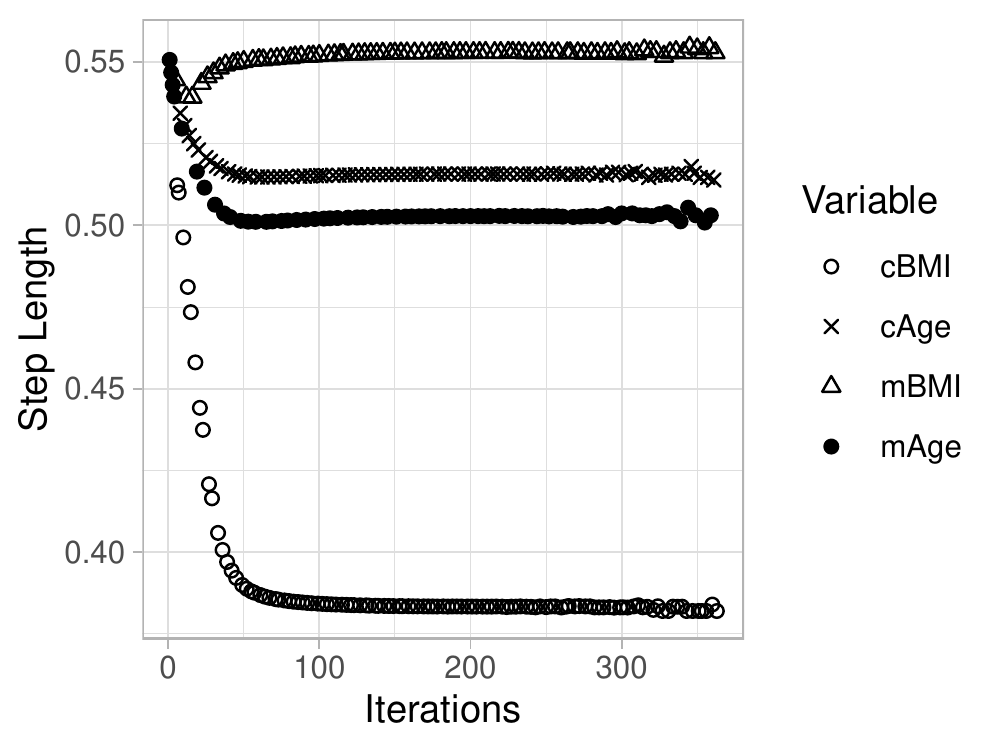}
\caption{$\nu_{j^*, \boldsymbol{\sigma}}^{*[m]}$} \label{fig: india_saasl_v_si}
\end{subfigure}
\caption{The optimal step-length of each model parameters against the boosting iterations. Up to the stopping iterations specified by 10-folds CV (here $m_{\text{stop}} = 769$), 406 iterations are used to update $\boldsymbol{\mu}$ and 363 iterations are used to update $\boldsymbol{\sigma}$.} \label{fig: india_saasl_v}
\end{figure}

%The reason why this example is not modeled well using either FSL or ASL (with a fixed search interval)  is the same as in the second simulation. 
%The large variance affects the empirical risk such that the negative partial derivatives of $\mu$ are very small when fitting the base-learners. 
%As a result, the algorithm updates $\mu$ very little in each iteration. 
Figure \ref{fig: india_saasl_v} presents the optimal step-lengths $\nu_{j^*, \boldsymbol{\mu}}^{*[m]}$ and $\nu_{j^*, \boldsymbol{\sigma}}^{*[m]}$ using SAASL for each variable up to 769 boosting iterations specified by 10-folds CV for one simulation run.
Apparently, the optimal step-lengths for $\boldsymbol{\mu}$ over the entire learning process are over 20000, which is far larger than the fixed step-length 0.1 and the upper boundary 10 of the predefined search interval in ASL. 
Without knowing this information, it is not trivial to determine the search interval for $\nu_{j^*, \boldsymbol{\mu}}^{*[m]}$.
And we thus (after acquiring this graphic) re-estimated the example data with ASL5.

Additionally, Figure \ref{fig: india_saasl_v}(\ref{fig: india_saasl_v_si}) illustrates the optimal step-length for $\boldsymbol{\sigma}$.
After several boosting iterations the optimal values of each covariate converge to their own stable regions (ranging from about 0.38 to 0.56).
As discussed above, the optimal step-lengths for $\boldsymbol{\sigma}$ should be some values around 0.5, and this graphic confirms this statement.

%In addition, the optimal step length plot in Fig.(\ref{fig: india_saasl_v_mu}) presents relationship between the coefficients of $\sigma$ and the optimal step length of $mu$. The estimated coefficients in Table \ref{tbl: india} of each variable present that $\beta_\text{cBMI} < \beta_\text{cAge} < \beta_\text{mAge} < \beta_\text{mBMI}$, the optimal step length fits also the same order after several boosting iterations.

%\begin{figure}[h]
%\centering
%\includegraphics[width=.8\linewidth]{./Figures/india_saasl_v_mu.pdf}
%\caption{The optimal step-length $\nu_{j^*, \boldsymbol{\mu}}^{*[m]}$ in SAASL}
%\label{fig: india_saasl_v_mu}
%\end{figure}

As this example  is not high-dimensional and does not necessarily require variable selection, we can use GAMLSS with penalized maximum likelihood estimation for comparison. 
The fact that its results are very similar to those of the semi-analytical approaches indicates  that  results from SAASL and SAASL05 are reliable. 
The only alternative to achieve balance between predictors would be using a
%The semi-analytical approaches are of course not the single solution to these kinds of data. Using the 
cyclical algorithm (with the downsides discussed in the introduction). 
%can also help to get an acceptable result. 
Rescaling the response variable or standardizing the negative partial derivatives could reduce the scaling problem to some extend, but would not eliminate the need to increase the step-length or reduce the imbalance between predictors. %But all of these methods except for the cyclical method require a deep understanding of the data and they are not as convenient and direct as SAASL.

\subsection{Riboflavin dataset}
\label{sec: ribo}
This data set describes the riboflavin (also known as vitamin $B_2$) production by Bacillus subtilis, containing 71 observations and 4088 predictors (gene expressions) [\cite{doi:10.1146/annurev-statistics-022513-115545, ribo}]. 
The log-transformed riboflavin production rate, which is close to a Gaussian distribution, is regarded as the response. 
This data set is chosen to demonstrate the capability of the boosting algorithm to deal with situations in which the number of covariates exceeds the number of observations. 
Please note that a comparison to the original GAMLSS algorithm is not possible in this case, since the algorithm is not able to deal with more model parameters than available observations.
In order to compare the out-of-sample MSE of each algorithm, we select 10 observations randomly as the validation set.
%Note that the number of covariates in this example is larger than the sample size and GAMLSS with penalized maximum likelihood estimation can thus not be used.
%Boosting is therefore the method of choice and allows for variable selection, but balanced updates for the models for mean and variance are important.

Table \ref{tbl: intersec_mu} summarize the selected informative variables for $\boldsymbol{\mu}$ and $\boldsymbol{\sigma}$ separately at the stopping iteration tuned by 5-fold CV, the corresponding coefficients are listed in Appendix \ref{apx: ribo}. 
The results in both tables demonstrate the intersection of the selected variables, for example FSL selects 13 informative variables in total, and 9 of them are also chosen by ASL and SAASL, and there are 11 variables common with SAASL05.
In general, for both $\boldsymbol{\mu}$ and $\boldsymbol{\sigma}$, more variables are included in the adaptive approaches and the difference in the selected variables mainly lies between the adaptive and fixed approach. 
Because the optimal step-length $\nu_{j^*, \boldsymbol{\mu}}^{*[m]}$ lies in the predefined search interval $[0, 10]$ (and is actually smaller than 1, i.e. the adaptive step-length $\nu_{j^*, \boldsymbol{\mu}}^{[m]} < 0.1$), and $\nu_{j^*, \boldsymbol{\sigma}}^{*[m]}$ lies also in a narrower predefined search interval $[0, 1]$, ASL and SAASL have the same results. 
Moreover, as the adaptive step-length is smaller than the fixed step-length 0.1, the adaptive approaches make  conservative (small) updates, leading to more  boosting iterations. 
Several of the gene expressions for $\boldsymbol{\mu}$ and $\boldsymbol{\sigma}$ are selected by all algorithms and are thus consistently included in the set of informative covariates.
Actually almost all gene expressions chosen by FSL are also recognized as informative variables by all other methods.

\begin{table}[h]
\centering
\caption{Number of chosen variables for $\eta_{\boldsymbol{\mu}}$ and $\eta_{\boldsymbol{\sigma}}$. The diagonal depicts the number per method, the off-diagonal elements overlapping variables.}
\label{tbl: intersec_mu}
\begin{tabular}{l|rrrr|rrrr}
\toprule
&	\multicolumn{4}{c|}{$\eta_{\boldsymbol{\mu}}$}	&	\multicolumn{4}{c}{$\eta_{\boldsymbol{\sigma}}$} \\
		&	FSL	&	ASL	&	SAASL	&	SAASL05	&	FSL	&	ASL	&	SAASL	&	SAASL05\\
\midrule
FSL		&	13	&	9	&	9		&	11		&	16	&	9	&	9	&	12\\
ASL		&	9	&	20	&	20		&	18		&	9	&	17	&	17	&	15\\
SAASL	&	9	&	20	&	20		&	18		&	9	&	17	&	17	&	15\\
SAASL05	&	11	&	18	&	18		&	24		&	12	&	15	&	15	&	24\\
\bottomrule
\end{tabular}
\end{table}

%
%\begin{table}[h]
%\centering
%\caption{Number of chosen variables for $\eta_{\boldsymbol{\sigma}}$. The diagonal depicts the number per method, the off-diagonal elements overlapping variables.}
%\label{tbl: intersec_si}
%\begin{tabular}{l|rrrr}
%\toprule
%{}		&	FSL	&	ASL	&	SAASL	&	SAASL05	\\
%\midrule
%FSL		&	16	&	9	&	9		&	12		\\
%ASL		&	9	&	17	&	17		&	15		\\
%SAASL	&	9	&	17	&	17		&	15		\\
%SAASL05	&	12	&	15	&	15		&	24		\\
%\bottomrule
%\end{tabular}
%\end{table}

To compare the performance of each algorithm, Table \ref{tbl: oos_mse}  lists the out-of-sample MSE. In contrast to the fixed approach, the three adaptive approaches perform in general well, where the performance of SAASL05 is slightly worse than the other two.
In addition, table \ref{tbl: oos_mse} demonstrates also the result of Lasso estimator from the R package \texttt{glmnet} [\cite{glmnet}] suggested by \cite{doi:10.1146/annurev-statistics-022513-115545}.
The mean squared prediction error of \texttt{glmnet} is the smallest among the five approaches, but the difference with the adaptive approaches is relatively small.

\begin{table}[h]
\centering
\caption{Out-of-sample MSE.} \label{tbl: oos_mse}
\begin{tabular}{l|rrrrr}
\toprule
&	FSL	&	ASL	&	SAASL	&	SAASL05 &	\texttt{glmnet}\\
\midrule
MSE	&	2.611	&	1.111	&	1.111	&	1.193	&	0.946\\
\bottomrule
\end{tabular}
\end{table}

As \texttt{glmnet} cannot model the scale parameter $\boldsymbol{\sigma}$, only the estimated coefficients of the $\boldsymbol{\mu}$-submodel are provided in Appendix \ref{apx: ribo}.
Out of the 21 genes selected by \texttt{glmnet}, 7 and 9 of them are common with the ASL/SAASL and SAASL05, respectively.
The signs (positive/negative) of the estimated coefficients of these common covariates from \texttt{glmnet} match the adaptive approaches.
This comparison indicates that the boosted GAMLSS with adaptive step-length is an applicable and competitive approach for high-dimensional data analysis.

\section{Conclusions and Outlook}
\label{sec: conclusion}
The step-length is often not treated as an important tuning parameter in many boosting algorithms, as long as it is set to a small value. 
However, if complex models like GAMLSS with several predictors for the different distribution parameters are estimated, the different scales of the distribution parameters can lead to imbalanced updates and resulting bad performances if one common small fixed step-length is used, as we show in this paper. 

The main contribution of this article is the proposal to use separate adaptive step-lengths for each distribution parameter in a non-cyclical boosting algorithm for GAMLSS,  which are optimized  in each iteration. 
In addition to the resulting balance in updates between different distribution parameters, a balance between over- and underfitting is obtained by taking only a proportion (shrinkage parameter) such as 10\% of the determined optimal step-length as the adaptive step-length. 
The optimal step-length can be found by optimization procedures such as a line search. 
We illustrated with an example the  importance of updating the search interval for the search if necessary to find the optimal solution. % but to find an appropriate searching interval sometimes needs multiple attempts.

%Just like the step length derived from the Gaussian distribution, there is a strong positive relationship between the adaptive step length of the location parameter and the variance of the response variable. That means if the variance of the response is very large, small step length will not make the estimation converge to the correct solution within limited boosting iterations. If we decide to increase the step length, the degree of step increase is the question that follows immediately. On the other hand, by comparing with the asymptotic adaptive step length 0.05 for the scale parameter, the common choice 0.1 is no longer small from our point of view.

For the important special case of the Gaussian distribution, we derived an analytical solution for the adaptive step-length for the mean parameter $\boldsymbol{\mu}$, which avoids numerical optimization and specification of a search interval. 
For the scale parameter $\boldsymbol{\sigma}$, we obtained an approximate solution of 0.5 (or 0.05 with 10\% proportion), which gives a better motivated default value than 0.1 relative to the step-length for $\boldsymbol{\mu}$, and discussed a combination with a one-dimensional line search in the semi-analytical approach.

In simulations and empirical applications, we showed favorable behavior compared to using a fixed step-length FSL. %Due to the adaptive step lengths in Gaussian distribution for scale parameter are more reasonable and usually smaller than 0.1, the false positive rates of adaptive approaches are lower than FSL. 
We showed highly competitive results of our adaptive approaches compared to a standard GAMLSS with respect to estimation accuracy for the low-dimensional case, while the adaptive boosting approach has the advantages of shrinkage and variable selection, which makes it also applicable to the high-dimensional case of more covariates than observations. 
Overall, the semi-analytical method for adaptive step-length selection performed  best among the considered  methods. 

In this paper we focused on the special case of the Gaussian distribution to derive analytical or semi-analytical solutions for the optimal step-length. 
In other cases, a line search has to be conducted for all distribution parameters. 
In the future, it is worth investigating to derive analytical adaptive step-lengths for other distributions as well, because analytical or approximate adaptive step-lengths increase the numerical efficiency and also reveal the relationships between the optimal step-lengths for different parameters and model parameters (as well as properties of commonly used but probably less than ideal step-length settings).

Further work should also be the implementation of further  (e.g. non-linear, spatial etc.) effects [\cite{doi:10.1890/10-0602.1}] into the model, and test the influence of the adaptive step-length on such effects.  %and the improvements on the performance of models. 
Moreover, we discovered correlations between the optimal step-length $\nu_{j^*, \boldsymbol{\mu}}^{*[m]}$ of a variable and the coefficient of this variable in the $\boldsymbol{\sigma}$-submodel through our application of the algorithm. 
Future work should also investigate the relationship among the  optimal step-lengths of different parameters and the relationship of these step-lengths to the model coefficients. 

%the relationship between the adaptive step length of the location parameter and the variance of the response in Gaussian distribution, Figure.(\ref{fig: india_saasl_v_mu}) demonstrates additionally the correlation between the optimal step length $\nu_\mu^*$ of a variable and the coefficient of this variable in $\sigma$ sub-model, i.e., after several iterations have been performed, if $\beta_{\sigma x_j}$ is large, then $\nu_{\mu (x_j)}$ is also large. A mathematical explanation for this correlation still needs to be completed.

A basic R package \texttt{ASL} based on this article is available online at https://github.com/FAUBZhang/ASL.
This package contains the source code of Algorithm \ref{alg: saasl} and the function of the corresponding cross-validation.
Some simple examples can also be found in this package.
This package is originated from the R package \texttt{gamboostLSS}, we hope to implement the functions of \texttt{ASL} into the latter in the future.

%Another problem deserving further research is the limitation of the estimation from boosting algorithms. If only the results of boosting are given, it is not possible to tell, with how much confidence we can trust the estimations, as with classical statistical models. Just as the malnutrition example, only when seeing the results of GAMLSS can we be less nervous about the SAASL estimations. But such comparison is not possible in riboflavin example, because GAMLSS cannot deal with high-dimensional data. We will further investigate the boosting framework, and discuss the possibility of establishing an estimator from which that has the statistical properties.

\subsection*{Acknowledgements}
The work on this article was supported by the Freigeist-Fellowships of Volkswagen Stiftung, project ``Bayesian Boosting - A new approach to data science, unifying two statistical philosophies". Boyao Zhang performed the present work in partial fulfilment of the requirements for obtaining the degree ``Dr. rer. biol. hum." at the Friedrich-Alexander-Universit\"at Erlangen-N\"urnberg (FAU).

\bibliographystyle{chicago}
\bibliography{reference}

\newpage
\appendix
\appendixpage
\section[Appendix A]{Derive the analytical ASL for the Gaussian distribution}
\label{apx: A}
Take the negative log-likelihood as the loss function, the loss for Gaussian distribution can be displayed as
\begin{align*}
\rho\left(\boldsymbol{y}, \{\eta_{\boldsymbol{\mu}}, \eta_{\boldsymbol{\sigma}}\}\right) =& -\log \left [\frac{1}{\left(\sqrt{2\pi}\right)^n} \det\left(\diag\left(\exp \left(-\eta_{\boldsymbol{\sigma}}(\boldsymbol{X})\right)\right)\right) \cdot \right. \\
&\left. \cdot \exp \left(-\frac{1}{2}(\boldsymbol{y} - \eta_{\boldsymbol{\mu}}\left(\boldsymbol{X})\right)^T \diag\left(\exp \left(-2\eta_{\boldsymbol{\sigma}}(\boldsymbol{X})\right)\right) (\boldsymbol{y} - \eta_{\boldsymbol{\mu}}(\boldsymbol{X})) \right) \right] \\
%\rho\left(\boldsymbol{y}, \{\boldsymbol{\eta}_\mu, \boldsymbol{\eta}_\sigma\}\right) &= -\log \left[\frac{1}{\sqrt{2 \pi}\exp(\boldsymbol{\eta}_\sigma)} \exp\left(-\frac{(\boldsymbol{y} - \boldsymbol{\eta}_\mu)^2}{2 \exp(2\eta_\sigma)}\right) \right] \\
=& \frac{n}{2} \log(2\pi) + \boldsymbol{1}_n^T \eta_{\boldsymbol{\sigma}}(\boldsymbol{X}) + \frac{1}{2}\left(\boldsymbol{y} - \eta_{\boldsymbol{\mu}}(\boldsymbol{X})\right)^T \diag\left(\exp\left(-2\eta_{\boldsymbol{\sigma}}(\boldsymbol{X})\right)\right)\left(\boldsymbol{y} - \eta_{\boldsymbol{\mu}}(\boldsymbol{X})\right).
%&= \frac{1}{2} \log(2\pi) + \boldsymbol{\eta}_\sigma + \frac{(y-\eta_\mu)^2}{2\exp(2\eta_\sigma)}.
\end{align*}
The negative partial derivatives for both distribution parameters in iteration $m$ are then
\begin{align}
\boldsymbol{u}_{\boldsymbol{\mu}}^{[m]} &= -\frac{\partial \rho\left(\boldsymbol{y}, \{\hat{\eta}_{\boldsymbol{\mu}}^{[m-1]}, \hat{\eta}_{\boldsymbol{\sigma}}^{[m-1]}\}\right)}{\partial \hat{\eta}_{\boldsymbol{\mu}}} \\
&= \diag\left(\exp\left(-2 \hat{\eta}_{\boldsymbol{\sigma}}^{[m-1]}(\boldsymbol{X})\right)\right) \left(\boldsymbol{y} - \hat{\eta}_{\boldsymbol{\mu}}^{[m-1]}(\boldsymbol{X})\right), \label{apx:eq: u_mu} \\ 
\boldsymbol{u}_{\boldsymbol{\sigma}}^{[m]} &= -\frac{\partial \rho\left(\boldsymbol{y}, \{\hat{\eta}_{\boldsymbol{\mu}}^{[m-1]}, \hat{\eta}_{\boldsymbol{\sigma}}^{[m-1]}\}\right)}{\partial \hat{\eta}_{\boldsymbol{\sigma}}} \\
&= -\boldsymbol{1}_n + \diag\left(\left(\boldsymbol{y} - \hat{\eta}_{\boldsymbol{\mu}}^{[m-1]}(\boldsymbol{X})\right)^T\right) \diag\left(\exp \left(-2 \hat{\eta}_{\boldsymbol{\sigma}}^{[m-1]}(\boldsymbol{X})\right)\right)\left(\boldsymbol{y} - \hat{\eta}_{\boldsymbol{\mu}}^{[m-1]}(\boldsymbol{X})\right). \label{apx:eq: u_si}
\end{align}
Both $\boldsymbol{u}_{\boldsymbol{\theta}}^{[m]}, \boldsymbol{\theta} \in \{\boldsymbol{\mu}, \boldsymbol{\sigma}\}$ can be regressed on the simple linear base-learner $h_{j^*, \boldsymbol{\theta}}^{[m]}(\boldsymbol{x}_{\cdot j^*})$, where $j^*$ denotes the best-fitting variable.
\begin{align}
\boldsymbol{u}_{\boldsymbol{\mu}}^{[m]} &= \hat{h}_{j^*, \boldsymbol{\mu}}^{[m]}(\boldsymbol{x}_{\cdot j^*}) + \hat{\boldsymbol{\epsilon}}_{\boldsymbol{\mu}}^{[m]} \label{apx: u_mu}\\
\boldsymbol{u}_{\boldsymbol{\sigma}}^{[m]} &= \hat{h}_{j^*, \boldsymbol{\sigma}}^{[m]}(\boldsymbol{x}_{\cdot j^*}) + \hat{\boldsymbol{\epsilon}}_{\boldsymbol{\sigma}}^{[m]},  \label{apx: u_si}
\end{align}
where $\hat{\boldsymbol{\epsilon}}_{\boldsymbol{\mu}}^{[m]}$ and $\hat{\boldsymbol{\epsilon}}_{\boldsymbol{\sigma}}^{[m]}$ denote the residuals in simple linear regression models.

\subsection{Optimal step-length for $\boldsymbol{\mu}$}
\label{apx: A_mu}
The analytical optimal step-length for $\boldsymbol{\mu}$ in iteration $m$ is obtained by minimizing the empirical risk,
\begin{align*}
\nu^{*[m]}_{j^*, \boldsymbol{\mu}} &= \argmin_{\nu} \sum_{i=1}^n \rho\left(y_i, \{\hat{\eta}_{\boldsymbol{\mu}}^{[m]}(\boldsymbol{x}_{i \cdot}), \hat{\eta}_{\boldsymbol{\sigma}}^{[m-1]}(\boldsymbol{x}_{i \cdot})\} \right) \\
&= \argmin_{\nu} \sum_{i=1}^n \rho\left(y_i, \{\hat{\eta}_{\boldsymbol{\mu}}^{[m-1]}(\boldsymbol{x}_{i\cdot}) + \nu \hat{h}_{j^*, \boldsymbol{\mu}}^{[m]}(x_{ij^*}), \hat{\eta}_{\boldsymbol{\sigma}}^{[m-1]}(\boldsymbol{x}_{i\cdot}) \} \right)\\
&= \argmin_{\nu} \sum_{i=1}^n -\log\left[\frac{1}{\sqrt{2\pi} \exp(\hat{\eta}_{\boldsymbol{\sigma}}^{[m-1]}(\boldsymbol{x}_{i\cdot}))} \exp\left(-\frac{\left(y_i - \hat{\eta}_{\boldsymbol{\mu}}^{[m-1]}(\boldsymbol{x}_{i\cdot}) - \nu\hat{h}_{j^*, \boldsymbol{\mu}}^{[m]}(x_{ij^*}) \right)^2}{2\exp(2\hat{\eta}_{\boldsymbol{\sigma}}^{[m-1]}(\boldsymbol{x}_{i\cdot}))} \right) \right] \\
&= \argmin_{\nu} \sum_{i=1}^n \left[\frac{1}{2}\log(2\pi) + \log(\hat{\sigma}_i^{[m-1]}) + \frac{\left(y_i - \hat{\eta}_{\boldsymbol{\mu}}^{[m-1]}(\boldsymbol{x}_{i\cdot}) - \nu \hat{h}_{j^*, \boldsymbol{\mu}}^{[m]}(x_{ij^*}) \right)^2}{2\sigma_i^{2[m-1]}} \right]\\
&= \argmin_{\nu} \sum_{i=1}^n \frac{\left(y_i - \hat{\eta}_{\boldsymbol{\mu}}^{[m-1]}(\boldsymbol{x}_{i\cdot}) - \nu \hat{h}_{j^*, \boldsymbol{\mu}}^{[m]}(x_{ij^*})\right)^2}{2\hat{\sigma}_i^{2[m-1]}},
\end{align*}
Note that the expression $\hat{\sigma}_i^{2[m-1]}$ represents the square of the standard deviation in the previous boosting iteration, i.e. $\hat{\sigma}_i^{2[m-1]} = (\hat{\sigma}_i^{[m-1]})^2$. 
And according to the model specification $\hat{\sigma}_i^{[m-1]} = \exp(\hat{\eta}_{\boldsymbol{\sigma}}^{[m-1]}(\boldsymbol{x}_{i\cdot}))$.

It can be shown, that the expression is a convex function, so the optimal value $\nu_{\boldsymbol{\mu}}^{*[m]}$ is accessed by letting the first order derivative equal zero,
\begin{align*}
& \frac{\partial}{\partial \nu} \sum_{i=1}^n \frac{\left(y_i - \hat{\eta}_{\boldsymbol{\mu}}^{[m-1]}(\boldsymbol{x}_{i\cdot}) - \nu \hat{h}_{j^*, \boldsymbol{\mu}}^{[m]}(x_{ij^*}) \right)^2}{2\hat{\sigma}_i^{2[m-1]}} \\
\overset{Eq.\eqref{apx:eq: u_mu}}{=}& \frac{\partial}{\partial \nu} \sum_{i=1}^n \frac{\left(u_{\boldsymbol{\mu}, i}^{[m]} \hat{\sigma}_i^{2[m-1]} -  \nu \hat{h}_{j^*, \boldsymbol{\mu}}^{[m]}(x_{ij^*})\right)^2}{2\hat{\sigma}_i^{2[m-1]}} \\
=& \frac{\partial}{\partial \nu} \sum_{i=1}^n \left(\frac{1}{2} u_{\boldsymbol{\mu}, i}^{2[m]} \hat{\sigma}_i^{2[m-1]} - \nu \hat{h}_{j^*, \boldsymbol{\mu}}^{[m]}(x_{ij^*}) u_{\boldsymbol{\mu}, i}^{[m]} + \frac{\nu^2 \left(\hat{h}_{j^*, \boldsymbol{\mu}}^{[m]}(x_{ij^*})\right)^2}{2\hat{\sigma}_i^{2[m-1]}} \right) \\
=& \sum_{i=1}^n \left(-\hat{h}_{j^*, \boldsymbol{\mu}}^{[m]}(x_{ij^*}) + \nu \frac{\left(\hat{h}_{j^*, \boldsymbol{\mu}}^{[m]}(x_{ij^*}) \right)^2}{\hat{\sigma}_i^{2[m-1]}} \right) \overset{!}{=} 0 \\
\Leftrightarrow  \nu &= \frac{\sum_{i=1}^n \hat{h}_{j^*, \boldsymbol{\mu}}^{[m]}(x_{ij^*}) u_{\boldsymbol{\mu}, i}^{[m]}}{\sum_{i=1}^n \frac{\left(\hat{h}_{j^*, \boldsymbol{\mu}}^{[m]}(x_{ij^*}) \right)^2}{\hat{\sigma}_i^{2[m-1]}}} \\
&\overset{Eq.\eqref{apx: u_mu}}{=} \frac{\sum_{i=1}^n \hat{h}_{j^*, \boldsymbol{\mu}}^{[m]}(x_{ij^*}) \left(\hat{h}_{j^*, \boldsymbol{\mu}}^{[m]}(x_{ij^*}) + \hat{\epsilon}_{\boldsymbol{\mu}, i}^{[m]} \right)}{\sum_{i=1}^n \frac{\left(\hat{h}_{j^*, \boldsymbol{\mu}}^{[m]}(x_{ij^*})\right)^2}{\hat{\sigma}_i^{2[m-1]}}} \\
&= \frac{\sum_{i=1}^n \left(\hat{h}_{j^*, \boldsymbol{\mu}}^{[m]}(x_{ij^*}) \right)^2 + \sum_{i=1}^n \hat{h}_{j^*, \boldsymbol{\mu}}^{[m]}(x_{ij^*}) \hat{\epsilon}_{\boldsymbol{\mu}, i}}{\sum_{i=1}^n \frac{\left(\hat{h}_{j^*, \boldsymbol{\mu}}^{[m]}(x_{ij^*})\right)^2}{\hat{\sigma}_i^{2[m-1]}}} \\
&= \frac{\sum_{i=1}^n \left(\hat{h}_{j^*, \boldsymbol{\mu}}^{[m]}(x_{ij^*}) \right)^2}{\sum_{i=1}^n \frac{\left(\hat{h}_{j^*, \boldsymbol{\mu}}^{[m]}(x_{ij^*})\right)^2}{\hat{\sigma}_i^{2[m-1]}}},
\end{align*}
where $\sum_{i=1}^n \hat{h}_{j^*\boldsymbol{\mu}}^{[m]}(x_{ij^*}) \hat{\epsilon}_{\boldsymbol{\mu}, i} = 0$, because the residuals are uncorrelated with the fitted values.

\subsection{Optimal step-length for $\boldsymbol{\sigma}$}
\label{apx: A_sigma}
The analytical optimal step-length for $\boldsymbol{\sigma}$ in iteration $m$ is obtained by minimizing the empirical risk,
\begin{align*}
\nu_{\boldsymbol{\sigma}}^{*[m]} =& \argmin_{\nu} \sum_{i=1}^n \rho\left(y_i, \{\hat{\eta}_{\boldsymbol{\mu}}^{[m-1]}(\boldsymbol{x}_{i\cdot}), \hat{\eta}_{\boldsymbol{\sigma}}^{[m]}(\boldsymbol{x}_{i\cdot})\} \right) \\
=& \argmin_{\nu} \sum_{i=1}^n \rho\left(y_i, \{\hat{\eta}_{\boldsymbol{\mu}}^{[m-1]}(\boldsymbol{x}_{i\cdot}), \hat{\eta}_{\boldsymbol{\sigma}}^{[m-1]}(\boldsymbol{x}_{i\cdot}) + \nu\hat{h}_{j^*, \boldsymbol{\sigma}}^{[m]}(x_{ij^*})\}\right) \\
=& \argmin_{\nu} \sum_{i=1}^n -\log\left[\frac{1}{\sqrt{2\pi} \exp\left(\hat{\eta}_{\boldsymbol{\sigma}}^{[m-1]}(\boldsymbol{x}_{i\cdot}) + \nu\hat{h}_{j^*, \boldsymbol{\sigma}}^{[m]}(x_{ij^*}) \right)} \cdot \right.\\
& \left. \cdot \exp\left(- \frac{\left(y_i - \hat{\eta}_{\boldsymbol{\mu}}^{[m-1]}(\boldsymbol{x}_{i\cdot}) \right)^2}{2 \exp\left(2\hat{\eta}_{\boldsymbol{\sigma}}^{[m-1]}(\boldsymbol{x}_{i\cdot}) + 2\nu \hat{h}_{j^*, \boldsymbol{\sigma}}^{[m]}(x_{ij^*}) \right)} \right) \right] \\
=& \argmin_{\nu} \sum_{i=1}^n \frac{1}{2}\log(2\pi) + \sum_{i=1}^n \left(\hat{\eta}_{\boldsymbol{\sigma}}^{[m-1]}(\boldsymbol{x}_{i\cdot}) + \nu\hat{h}_{j^*, \boldsymbol{\sigma}}^{[m]}(x_{ij^*}) \right) + \\
& \sum_{i=1}^n \frac{\left(y_i - \hat{\eta}_{\boldsymbol{\mu}}^{[m-1]}(\boldsymbol{x}_{i\cdot}) \right)^2}{2\exp\left(2\hat{\eta}_{\boldsymbol{\sigma}}^{[m-1]}(\boldsymbol{x}_{i\cdot}) + 2\nu\hat{h}_{j^*, \boldsymbol{\sigma}}^{[m]}(x_{ij^*})\right)} \\
=& \argmin_{\nu} \sum_{i=1}^n \left(\hat{\eta}_{\boldsymbol{\sigma}}^{[m-1]}(\boldsymbol{x}_{i\cdot}) + \nu\hat{h}_{j^*, \boldsymbol{\sigma}}^{[m]}(x_{ij^*}) \right) + \sum_{i=1}^n \frac{\left(y_i - \hat{\eta}_{\boldsymbol{\mu}}^{[m-1]}(\boldsymbol{x}_{i\cdot}) \right)^2}{2\exp\left(2\hat{\eta}_{\boldsymbol{\sigma}}^{[m-1]}(\boldsymbol{x}_{i\cdot}) + 2\nu\hat{h}_{j^*, \boldsymbol{\sigma}}^{[m]}(x_{ij^*})\right)}.
\end{align*}
It can be shown, that the second order derivative of the expression is positive and thus the expression a convex function.
Letting the first order derivative equal zero, we get
\begin{align*}
& \frac{\partial}{\partial \nu} \left[\sum_{i=1}^n \left(\hat{\eta}_{\boldsymbol{\sigma}}^{[m-1]}(\boldsymbol{x}_{i\cdot}) + \nu\hat{h}_{j^*, \boldsymbol{\sigma}}^{[m]}(x_{ij^*}) \right) + \sum_{i=1}^n \frac{\left(y_i - \hat{\eta}_{\boldsymbol{\mu}}^{[m-1]}(\boldsymbol{x}_{i\cdot}) \right)^2}{2\exp\left(2\hat{\eta}_{\boldsymbol{\sigma}}^{[m-1]}(\boldsymbol{x}_{i\cdot}) + 2\nu\hat{h}_{j^*, \boldsymbol{\sigma}}^{[m]}(x_{ij^*})\right)} \right] \\
=& \sum_{i=1}^n \hat{h}_{j^*, \boldsymbol{\sigma}}^{[m]}(x_{ij^*}) - \sum_{i=1}^n \left(y_i - \hat{\eta}_{\boldsymbol{\mu}}^{[m-1]}(\boldsymbol{x}_{i\cdot}) \right)^2 \hat{h}_{j^*, \boldsymbol{\sigma}}^{[m]}(x_{ij^*}) \exp \left(-2\hat{\eta}_{\boldsymbol{\sigma}}^{[m-1]}(\boldsymbol{x}_{i\cdot}) - 2\nu \hat{h}_{\boldsymbol{\sigma}}^{[m]}(x_{ij^*}) \right) \\
\overset{Eq.\eqref{apx:eq: u_si}}{=}& \sum_{i=1}^n \hat{h}_{j^*, \boldsymbol{\sigma}}^{[m]}(x_{ij^*}) - \sum_{i=1}^n \frac{u_{\boldsymbol{\sigma}, i}^{[m]} + 1}{\exp \left(-2\hat{\eta}_{\boldsymbol{\sigma}}^{[m-1]}(\boldsymbol{x}_{i\cdot}) \right)} \hat{h}_{j^*, \boldsymbol{\sigma}}^{[m]}(x_{ij^*}) \exp \left(-2\hat{\eta}_{\boldsymbol{\sigma}}^{[m-1]}(\boldsymbol{x}_{i\cdot}) - 2\nu \hat{h}_{j^*, \boldsymbol{\sigma}}^{[m]}(x_{ij^*}) \right) \\
=& \sum_{i=1}^n \hat{h}_{j^*, \boldsymbol{\sigma}}^{[m]}(x_{ij^*}) - \sum_{i=1}^n \left(u_{\boldsymbol{\sigma}, i}^{[m]} + 1 \right) \hat{h}_{j^*, \boldsymbol{\sigma}}^{[m]}(x_{ij^*}) \exp\left(- 2\nu \hat{h}_{j^*, \boldsymbol{\sigma}}^{[m]}(x_{ij^*}) \right) \\
\overset{Eq.\eqref{apx: u_si}}{=}& \sum_{i=1}^n \hat{h}_{j^*, \boldsymbol{\sigma}}^{[m]}(x_{ij^*}) - \sum_{i=1}^n \frac{\left(\hat{h}_{j^*, \boldsymbol{\sigma}}^{[m]}(x_{ij^*}) + \hat{\epsilon}_{\boldsymbol{\sigma}, i}^{[m]} + 1 \right) \hat{h}_{j^*, \boldsymbol{\sigma}}^{[m]}(x_{ij^*})}{\exp\left(2\nu\hat{h}_{j^*, \boldsymbol{\sigma}}^{[m]}(x_{ij^*}) \right)} \overset{!}{=} 0
\end{align*}

\newpage
\section[Appendix B]{Further graphics of section \ref{sec: sim1}}
\begin{figure}[H]
\centering
\includegraphics[width=\linewidth]{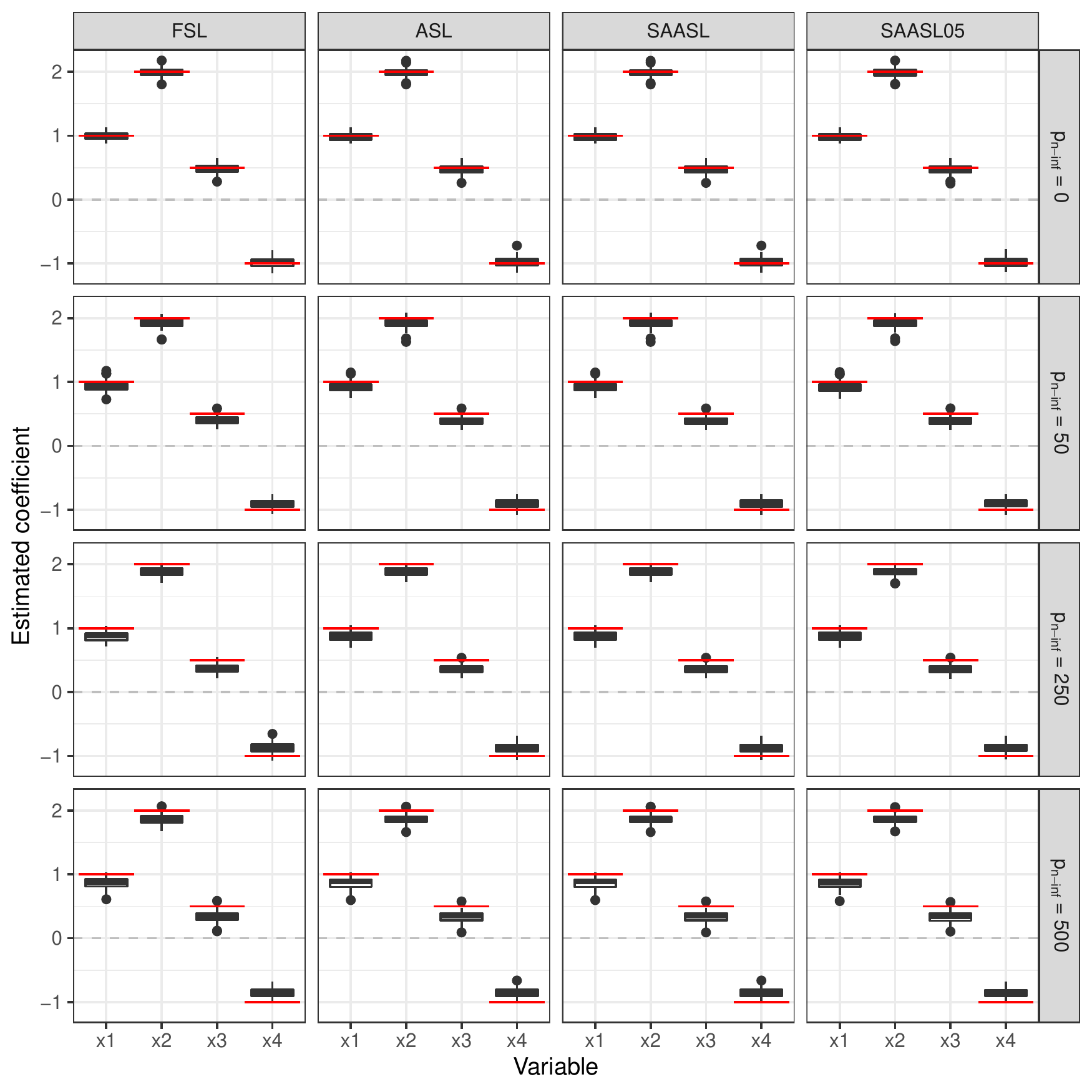}
\caption{Boxplot of the estimated coefficients of $\eta_\mu$ in 100 simulation runs. Values are taken at the stopping iterations determined by 10-folds cross-validation. The results are separated according to fixed and adaptive approaches with respect to different non-informative variables settings, i.e. $p_{\text{n-inf}} = 0, 50, 250$ and $500$. The horizontal red lines indicate the true coefficients. The shrinkage of the coefficients towards zero can be observed from this graphic.}
\label{fig: sim1_muMat}
\end{figure}

\begin{figure}[H]
\centering
\includegraphics[width=\linewidth]{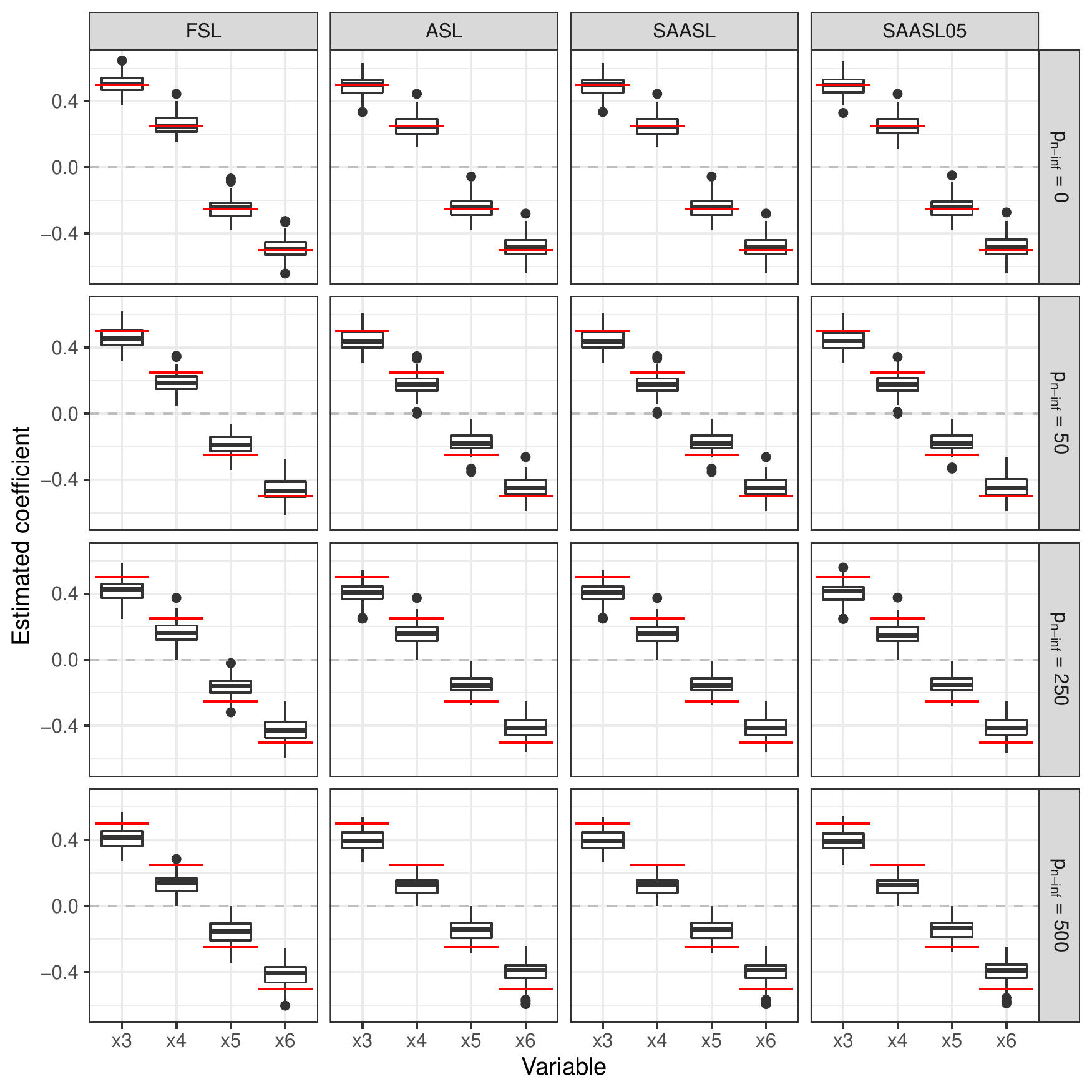}
\caption{Boxplot of the estimated coefficients of $\eta_\sigma$ in 100 simulation runs. Values are taken at the stopping iterations tuned by 10-folds cross-validation. The results are separated according to fixed and adaptive approaches with respect to different non-informative variables settings, i.e. $p_{\text{n-inf}} = 0, 50, 250$ and $500$. The horizontal red lines indicate the true coefficients. The shrinkage of the coefficients towards zero can be observed from this graphic.}
\label{fig: sim1_siMat}
\end{figure}

\begin{figure}[H]
\centering
\includegraphics[scale=.6]{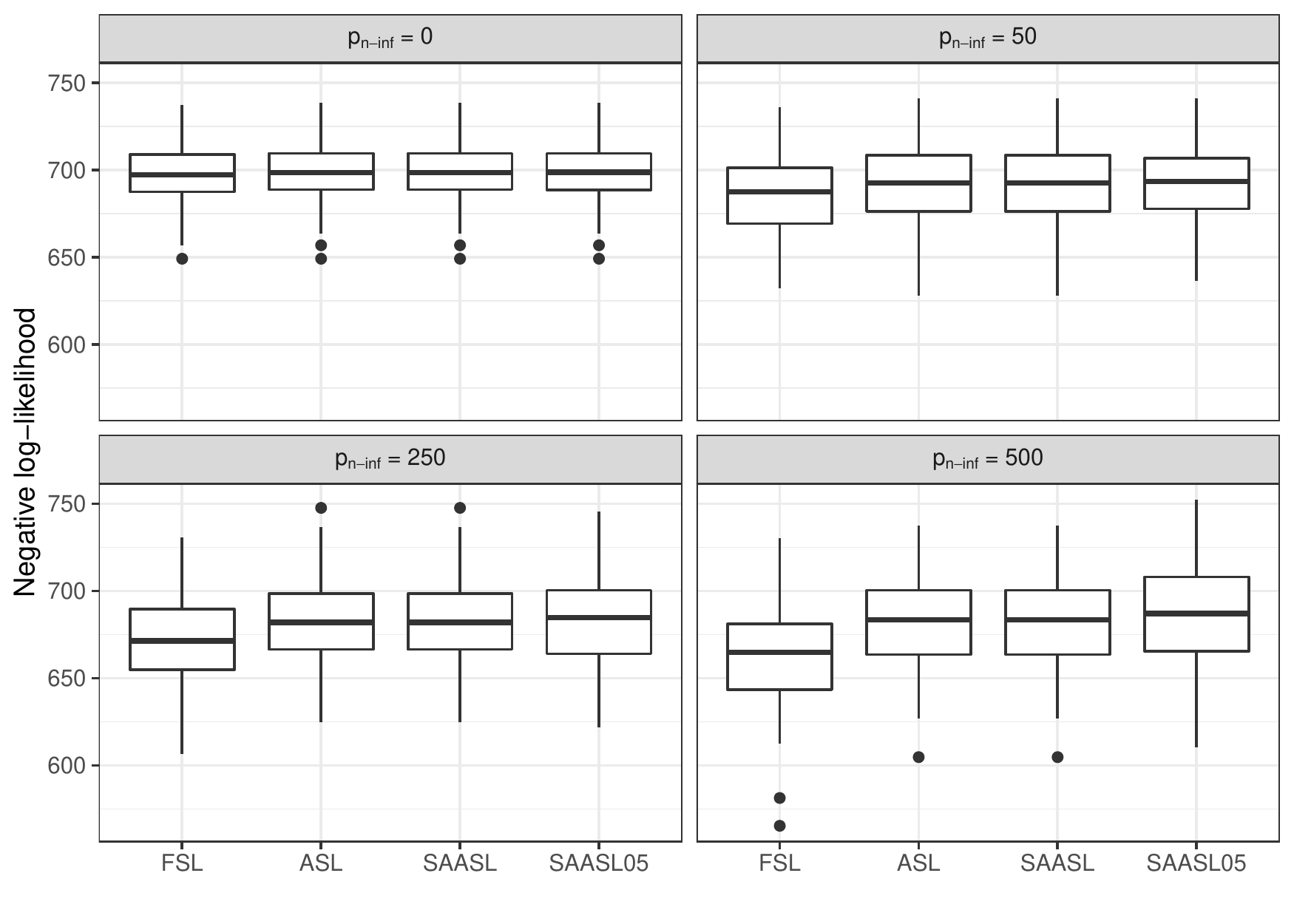}
\caption{Summary of the negative log-likelihood of 100 simulation runs with different estimating approaches with respect to various non-informative variables settings. Values are taken at the stopping iteration determined by 10-folds cross-validation.}
\label{fig: sim1_like}
\end{figure}

\begin{figure}[H]
\centering
\includegraphics[scale=.6]{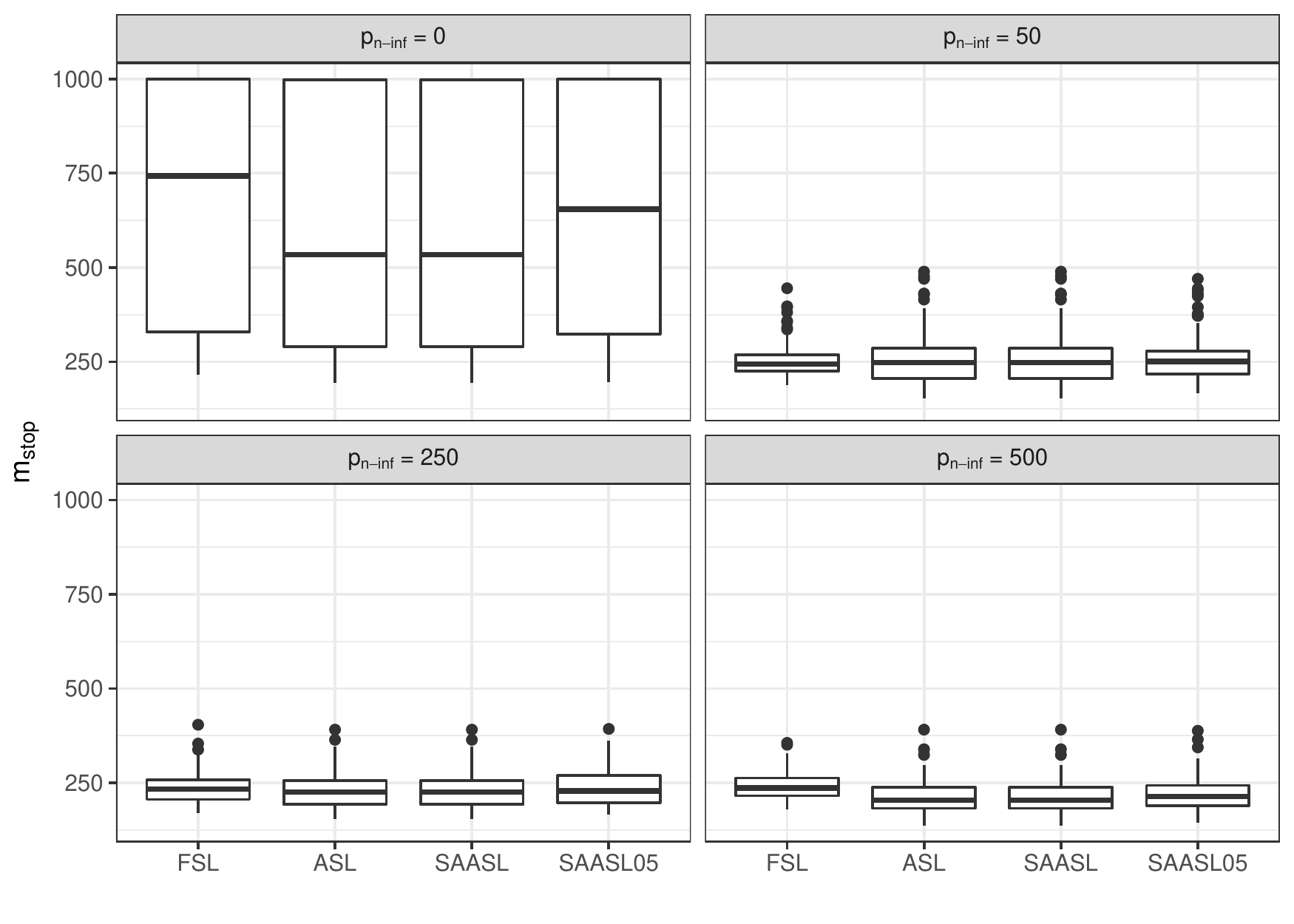}
\caption{$m_{\text{stop}}$ tuned by 10-fold CV with different estimating methods with respect to different non-informative variables settings. The predefined maximal learning iteration is 1000.}
\end{figure}

\section[Appendix C]{Further table of section \ref{sec: sim2}}
\label{apx: mse_table}
\begin{table}[H]
\centering
\caption{The average MSE of the estimated coefficients for both model parameters $\boldsymbol{\mu}$ and $\boldsymbol{\sigma}$ w.r.t. three estimation approaches. The MSE for each coefficient is calculated not only from 100 simulation runs (Total) at their stopping iterations but also from the true positive subsets (TP), i.e., the simulations from which a coefficient is selected by all three approaches.} %\label{tbl: sim2_coef_mse}
\begin{tabular}{c|c|c|c|c|c|c|c|c|c|c|c|c}
\toprule
		&	\multicolumn{6}{c|}{$\boldsymbol{\mu}$} 	&	\multicolumn{6}{c}{$\boldsymbol{\sigma}$} \\
		\hline
		&	\multicolumn{2}{c|}{$\hat{\beta}_1$}	&	\multicolumn{2}{c|}{$\hat{\beta}_2$}	&	\multicolumn{2}{c|}{$\hat{\beta}_3$}	&	\multicolumn{2}{c|}{$\hat{\beta}_1$}	&	\multicolumn{2}{c|}{$\hat{\beta}_2$}	&	\multicolumn{2}{c}{$\hat{\beta}_3$}\\
		\hline
		&	Total	&	TP	&	Total	&	TP	&	Total	&	TP	&	Total	&	TP	&	Total	&	TP	&	Total	&	TP\\
\hline
FSL		&	13.7	&	80.9	&	26.5	&	112.2	&	13.8	&	76.0	&	0.84	&	0.81	&	1.41	&	1.42	&	0.84	&	0.82\\
GAMLSS	&	113.8	&	328.8	&	145.6	&	355.8	&	116.8	&	339.4	&	0.82 	&	0.79	&	1.44	&	1.44	&	0.81	&	0.79\\
SAASL	&	71.3	&	250.2	&	95.8	&	271.7	&	73.3	&	238.5	&	0.85	&	0.81	&	1.39	&	1.40	&	0.85	&	0.82\\
\bottomrule
\end{tabular}
\end{table}

\section[Appendix D]{Estimated coefficients of riboflavin dataset}
\label{apx: ribo}
\begin{table}[H]
\centering
\caption{The estimated coefficients of the $\boldsymbol{\mu}$-submodel with fixed and adaptive approaches. Values are taken at the $m_{\text{stop}}$ tuned by 5-folds CV.} \label{tbl: coef_mu}
\begin{tabular}{rlrrrrr}
  \toprule
 & Variable & FSL & ASL & SAASL & SAASL05 & glmnet \\ 
  \midrule
1 & (Intercept) & -7.03 & -7.04 & -7.04 & -7.03 & 0.72 \\ 
  2 & ARGF\_at & -0.08 & -0.02 & -0.02 & -0.02 &  \\ 
  3 & IOLE\_at & -0.32 &  &  & -0.02 &  \\ 
  4 & LYSC\_at &  &  &  &  & -0.06 \\ 
  5 & RPLO\_at &  &  &  & -0.02 & -0.08 \\ 
  6 & SPOIISA\_at & 0.35 & 0.19 & 0.19 & 0.23 & 0.02 \\ 
  7 & XKDC\_at & 0.18 &  &  & 0.09 & 0.19 \\ 
  8 & XKDO\_at &  & 0.02 & 0.02 & 0.03 &  \\ 
  9 & XKDS\_at & 0.11 & 0.08 & 0.08 & 0.08 & 0.06 \\ 
  10 & XLYA\_at &  &  &  &  & 0.05 \\ 
  11 & XTMA\_at &  & 0.03 & 0.03 & 0.01 &  \\ 
  12 & XTRA\_at &  &  &  &  & 0.01 \\ 
  13 & YCDH\_at &  &  &  &  & -0.03 \\ 
  14 & YCGM\_at & -0.09 & -0.06 & -0.06 & -0.06 & -0.01 \\ 
  15 & YCGN\_at &  & -0.04 & -0.04 & -0.04 &  \\ 
  16 & YCGO\_at & -0.07 &  &  &  & -0.14 \\ 
  17 & YCGP\_at &  &  &  & -0.03 &  \\ 
  18 & YCKE\_at & 0.15 & 0.12 & 0.12 & 0.14 & 0.15 \\ 
  19 & YCLB\_at & 0.29 &  &  &  &  \\ 
  20 & YCSG\_at &  & -0.08 & -0.08 & -0.18 &  \\ 
  21 & YDAO\_at &  &  &  & -0.03 &  \\ 
  22 & YDAR\_at & -0.24 & -0.16 & -0.16 & -0.16 &  \\ 
  23 & YDDK\_at &  &  &  &  & -0.04 \\ 
  24 & YEBC\_at &  &  &  &  & -0.55 \\ 
  25 & YHAI\_at &  & 0.12 & 0.12 & 0.11 &  \\ 
  26 & YHFU\_at &  & -0.02 & -0.02 & -0.03 & -0.01 \\ 
  27 & YJCJ\_at & 0.11 & 0.04 & 0.04 & 0.04 &  \\ 
  28 & YKBA\_at &  &  &  &  & 0.01 \\ 
  29 & YKUH\_at &  & 0.05 & 0.05 & 0.06 &  \\ 
  30 & YOAB\_at &  &  &  &  & -0.34 \\ 
  31 & YORB\_i\_at &  & 0.03 & 0.03 & 0.05 & 0.10 \\ 
  32 & YOZH\_i\_at &  & 0.02 & 0.02 &  &  \\ 
  33 & YPGA\_at &  &  &  &  & -0.05 \\ 
  34 & YTGB\_at &  &  &  &  & -0.09 \\ 
  35 & YWQD\_at &  &  &  & -0.02 &  \\ 
  36 & YXJA\_at &  & -0.01 & -0.01 & -0.01 &  \\ 
  37 & YXLC\_at &  & -0.03 & -0.03 &  &  \\ 
  38 & YXLD\_at & -0.12 & -0.14 & -0.14 & -0.16 & -0.14 \\ 
  39 & YXLE\_at & -0.06 & -0.01 & -0.01 & -0.01 &  \\ 
   \bottomrule
\end{tabular}
\end{table}

\begin{table}
\centering
\caption{The estimated coefficients of $\boldsymbol{\sigma}$-submodel with fixed and adaptive approaches. Values are taken at the $m_{\text{stop}}$ tuned by 5-folds CV.} \label{tbl: coef_si}
\begin{tabular}{rlrrrr}
  \toprule
 & Variables & FSL & ASL & SAASL & SAASL05 \\ 
  \midrule
1 & (Intercept) & -1.41 & -1.29 & -1.29 & -1.53 \\ 
  2 & COTJC\_at &  & -0.18 & -0.18 &  \\ 
  3 & DEGA\_at & -0.13 & -0.61 & -0.61 & -0.89 \\ 
  4 & EXPZ\_at &  & 0.21 & 0.21 & 0.05 \\ 
  5 & LEVD\_at & 0.24 & 0.15 & 0.15 & 0.19 \\ 
  6 & NTH\_at & -0.09 & -0.11 & -0.11 &  \\ 
  7 & PHRI\_r\_at & 0.06 &  &  & 0.03 \\ 
  8 & TRUA\_at & -0.09 & -0.74 & -0.74 & -0.71 \\ 
  9 & XLYA\_at & -0.06 &  &  &  \\ 
  10 & XPF\_at &  &  &  & -0.05 \\ 
  11 & YACN\_at & 0.20 &  &  &  \\ 
  12 & YCNK\_at & 0.66 & 0.06 & 0.06 & 0.27 \\ 
  13 & YFIG\_at & -0.08 &  &  & -0.09 \\ 
  14 & YFMD\_at & -0.25 & -0.43 & -0.43 & -0.35 \\ 
  15 & YHBD\_at &  & -0.21 & -0.21 & -0.19 \\ 
  16 & YHEN\_at &  &  &  & -0.05 \\ 
  17 & YHFS\_at &  &  &  & 0.06 \\ 
  18 & YITQ\_at &  &  &  & -0.24 \\ 
  19 & YJFB\_at &  & -0.11 & -0.11 & -0.09 \\ 
  20 & YKRS\_at &  & 0.29 & 0.29 & 0.35 \\ 
  21 & YKVV\_at &  & 0.06 & 0.06 & 0.28 \\ 
  22 & YPGA\_at & 0.07 & 0.05 & 0.05 & 0.12 \\ 
  23 & YSBA\_at &  & -0.55 & -0.55 & -0.28 \\ 
  24 & YSBB\_at & -0.15 & -0.23 & -0.23 & -0.20 \\ 
  25 & YTFP\_at & -0.24 &  &  &  \\ 
  26 & YTQI\_at & 0.10 & 0.16 & 0.16 & 0.11 \\ 
  27 & YURR\_at &  & -0.07 & -0.07 & -0.11 \\ 
  28 & YWQA\_at & -0.11 &  &  & -0.20 \\ 
  29 & YYAE\_at &  &  &  & -0.06 \\ 
  30 & YYBT\_at & -0.08 &  &  & -0.03 \\ 
   \bottomrule
\end{tabular}
\end{table}

\end{document}